\documentclass[preprint2]{aastex7}
\usepackage{hyperref}
\usepackage{aas_macros}
\usepackage{textcomp}
\usepackage[T1]{fontenc}
\usepackage[utf8]{inputenc}
\received{May 14, 2026}
\accepted{September 16, 2026}

\begin{document}

\title{What Drives Galaxy-Galaxy Differences In The Selective Attenuation Curve? \\New Perspectives from DESI DR1}

\author[sname='Andrew Mizener']{Andrew Mizener}
\affiliation{University of Massachusetts Amherst, 710 North Pleasant Street, Amherst, MA 01003, USA}
\email[show]{amizener@umass.edu}

\author[sname='Daniela Calzetti']{Daniela Calzetti}
\affiliation{University of Massachusetts Amherst, 710 North Pleasant Street, Amherst, MA 01003, USA}
\email[hide]{calzetti@umass.edu}  

\author[sname='Ranga-Ram Chary']{Ranga-Ram Chary}
\affiliation{University of California, Los Angeles, CA 90095-1562, USA}
\email[hide]{rchary@g.ucla.edu} 

\author[sname='John Moustakas']{John Moustakas}
\affiliation{Siena University, 515 Loudon Road, Loudonville, NY 12211, USA}
\email[hide]{jmoustakas@siena.edu} 

\begin{abstract}

Leveraging a sample of 3457 galaxies drawn from DESI DR1 and cross-matched with archival GALEX UV photometry, we construct new dust attenuation curves and measure how the shape of these curves vary as a function of galaxy properties. Our curves are built by comparing the median flux densities of galaxies that have been divided into bins of nebular opacity, measured via the Balmer optical depth (i.e. the H$\alpha$/H$\beta$ ratio). When no further cuts on galaxy properties are applied, we find an attenuation curve very similar to previous work at low redshifts. We then test how the attenuation curves vary as a function of galaxy properties; stellar mass, specific star formation rate, metallicity, and ionization parameter. We find that the curve steepens significantly at higher specific star formation rate. In our data, there is not a clear link between the ionization parameter, mass, or metallicity and the attenuation curve slope. 

\end{abstract}

\keywords{\uat{Galaxies}{573} --- \uat{Interstellar medium}{847} --- \uat{Interstellar dust extinction}{837} --- \uat{Galaxy properties}{615}}


\section{Introduction} 
Dust is a universal presence within galaxies and has a major impact on their observed SEDs by reddening and attenuating light \citep{Salim2020}. The wavelength-dependent nature of this phenomenon is encoded by the attenuation curve \citep{Calzetti94}. Measuring attenuation curve shapes has long been of interest, as one must be assumed in order to correct for the presence of dust and measure the intrinsic properties of any galaxy. In recent years, it has become increasingly clear that the shape of these curves varies significantly from galaxy to galaxy -- particularly in the rest-frame ultraviolet \citep[e.g.][]{Salim18,Shivaei25}. This variation includes both the overall slope in the UV as well as the strength of the 2175 $\mathrm{\AA}$ feature. Depending on exactly what measurements are available, derived galaxy properties such as the star formation rate and stellar mass may be sensitive to attenuation assumptions \citep{Salim2020}. Even photometric redshifts might be biased by dust \citep{dust_photoz,eazy}, a particular concern for projects such as Roman and Euclid which require precisely calibrated photo-zs (on the order of $\sim 1$ to 5 $\%$) in order to derive reliable cosmological parameters \citep{Amara2007,Euclid2024,RomanScienceTeam2023}. It is thus important that we constrain which galaxy properties are correlated with variation in the shape of the attenuation curve.

Both extinction and attenuation curves have shapes driven by the size, shape, and composition of dust grains \citep{Gordon23}. In contrast to extinction curves, which only contain information about dust grain properties, attenuation curves are \textit{also} strongly affected by the spatial distribution of dust relative to sources of emission \citep{wittgordon,calzetti,Calzetti94}. Relevant geometric effects may include scattering into and out of the line of sight, differential extinction between stellar populations of varying age, and radiative transfer effects such as the anisotropy of scattering as a function of color \citep{Shivaei25,Chev13}. Of these, scattering is probably the most important in terms of its effect on curve shapes \citep{Matsumoto26}. Attenuation curve slopes are generally correlated with galaxy opacity, with more opaque galaxies showing shallower curves \citep{Salim18,Shivaei25}; however, there is substantial scatter in this relationship and other galaxy properties certainly contribute \citep{Salim18,shivaei2020}. 

The techniques used to derive the shape of the attenuation curve can be divided into two categories: one family of methods that compare galaxy SEDs empirically (the "pairwise method";  \cite{Calzetti94,shivaei2020,Battisti2016}) and another based on SED modeling (the "model method"; \cite{Salim18,Shivaei25}). The pairwise method, used in this work, is nonparametric but cannot be used to derive curves for individual galaxies and implicitly assumes that every attenuation curve within a given sub-sample is identical. Additionally, this method uses nebular attenuation as a proxy for the stellar continuum attenuation which may introduce scatter \citep{Salim18}. While model-based methods can infer the curves for individual galaxies, they are parametric by nature and may be degenerate with the characteristics of the input stellar population models and underlying star formation history (SFH). As a result, both methods have a place in our contemporary effort to understand the attenuation curve.

Our paper expands on the results of \citet{Battisti2016}, which used a similar methodology our work, by extending the redshift out to z=0.25 (versus z=0.1 in the that study) and by targeting a distinct galaxy population at lower metallicity (with Log(O/H)+12 from 8.1-8.6 for the bulk of galaxies in our work compared to Log(O/H)+12 $\sim$ 8.5-9.2 there.), a likely consequence of the cuts in angular size for our sample. We also study the attenuation curves as a function of metallicity, mass, and ionization parameter, which was not attempted in Battisti et al. (2016) due to cuts on these parameters producing samples with a limited dynamic range in $\tau_b$. We use a new strategy to bin galaxies in $\tau_b$, which allows us to produce curves despite this issue. We also explore the effects of redshift, concluding that there are no significant redshift effects on the attenuation curve. (Appendix C).

This paper is structured as follows. In Section 2, we present the data used to derive our attenuation curves and describe the various quality cuts applied to the catalogs. Section 3 describes the methods we used to process these data and produce our attenuation curves. The properties of our sample, and the attenuation curves we derive, are shown in Section 4. Finally, we discuss the implications of our results in Section 5. Section 6 summarizes the work as a whole. Throughout this work, we assume a Chabrier IMF \citep{Chab} and a standard Planck 2018 (flat $\Lambda$CDM) cosmology \citep{Planck18}; $H_0$ = 67.7, $\Omega_{M}$ = 0.310.

\section{Data}

\subsection{What Makes An Attenuation Curve?}
In order to produce attenuation curves using the pairwise method, constraints on the stellar continuum shape and attenuation of each galaxy are required. Constraints on attenuation may be obtained by measuring the Balmer optical depth $\tau_{b}$ \citep{Calzetti94}. This is derived from the ratio of the observed $F_{\mathrm{H}\alpha} / F_{\mathrm{H}\beta}$ to the theoretical lower limit of this value assuming Case B recombination, usually assumed to be 2.86 \citep{Osterbrock}, such that 
\begin{equation} \tau_{b} =  \ln(\frac{F(H\alpha)/F(H\beta)}{2.86}) \end{equation} 
Stellar continuum reddening is often understood through the UV slope $\beta$, which in the context of GALEX data is typically defined as \begin{equation} \beta =  \frac{\log(F_{\lambda}(FUV)/F_{\lambda}(NUV))}{\log(\lambda_{FUV}/\lambda_{NUV})} \end{equation} 
In general, the minimum data required to measure the attenuation curve via the pairwise method is two Balmer (or other Hydrogen recombination) lines, 2 measurements in the UV (to constrain the UV slope), and sufficient information in the rest-optical to constrain the nebular-to-stellar differential reddening (optimally corresponding to measurements in the rest-frame B and V bands).
Although nebular reddening and stellar continuum reddening are distinct phenomena, the two are correlated (typical Spearman $\rho_s$ values cluster around $\sim$0.5, though this varies from study to study) allowing for the Balmer decrement to be used as a tracer for either \citep[e.g.][]{Calzetti94,Battisti2016,reddy15}. 

Any deviation from the Case B line ratio is assumed to be the result of attenuation by dust, as the (bluer) $\mathrm{H}\beta$ line will be subject to a higher optical depth than the (redder) $\mathrm{H}\alpha$ line for the same amount of total attenuation. The assumption that the physical lower bound on the $F_{\mathrm{H}\alpha} / F_{\mathrm{H}\beta}$ is 2.86 is generally reasonable, though the true Case B line ratio will vary somewhat as a function of electron temperature. Typical variation in this ratio is around $\sim11\%$ \citep{Osterbrock}. Though some extreme galaxies have intrinsic $F_{\mathrm{H}\alpha} / F_{\mathrm{H}\beta}$ line ratios that deviate strongly from Case B \citep{Scarlata24}, this is rare. Small deviations from Case B are more likely to result from uncertain line measurements.

\subsection{DESI-GALEX Selection And Matching}
In this work, we use data from the Dark Energy Spectroscopic Instrument (DESI) to measure $\tau_{b}$ and constrain the shape of the stellar continuum. We derive $\tau_{b}$ from the DESI Data Release 1 (DR1) \texttt{IRON} Value Added Catalogue (VAC) version 2.1 (\texttt{FastSpecFit}) $\mathrm{H}\alpha$ and $\mathrm{H}\beta$ \citep{DESI,fsf,Mous}. We use GALEX FUV and NUV photometry \citep{Bianchi14} to constrain the UV continuum shape of our galaxies. We also make use of \texttt{TRACTOR}  \textit{grz} morphological fits from the Legacy Surveys Data Release 9 \citep{legsurvey}. A summary of the cuts we apply to our sample are given in Table 1. The effects of these cuts on our sample are described in Appendix A.

\subsubsection{DESI Quality Cuts}

We begin by applying a number of quality cuts based on the DESI \texttt{IRON} catalog. First, some galaxies are observed multiple times in DESI. We remove these duplicates by only using the observations associated with the \texttt{ZCAT\_PRIMARY = TRUE} flag. This flag is only set \texttt{TRUE} once per unique target and denotes the best available observation for that target.

Next, we select galaxies with a $SNR > 3$ on the 2 relevant Balmer lines, $\mathrm{H}\alpha$ and $\mathrm{H}\beta$. This SNR cut is performed on the line amplitudes rather than the line fluxes, as the catalog uncertainty on the line fluxes is not propagated correctly and so results in an overestimated line SNR for many galaxies. This is a more conservative cut and preferentially removes galaxies between about $z\sim0.14$ to $0.18$ (Figure 1). This is related to the spectral coverage of the DESI instrument; for the affected galaxies, $\mathrm{H}\alpha$ falls in the overlap region between the Red and NIR channels, while $\mathrm{H}\beta$ falls in the overlap region between the Blue and Red channels \citep{DESI,guy23}. We further restrict our sample to galaxies corresponding to a redshift of 0.25 or less.

\begin{figure}[htb!]
\centering
\includegraphics[width=1.0\columnwidth]{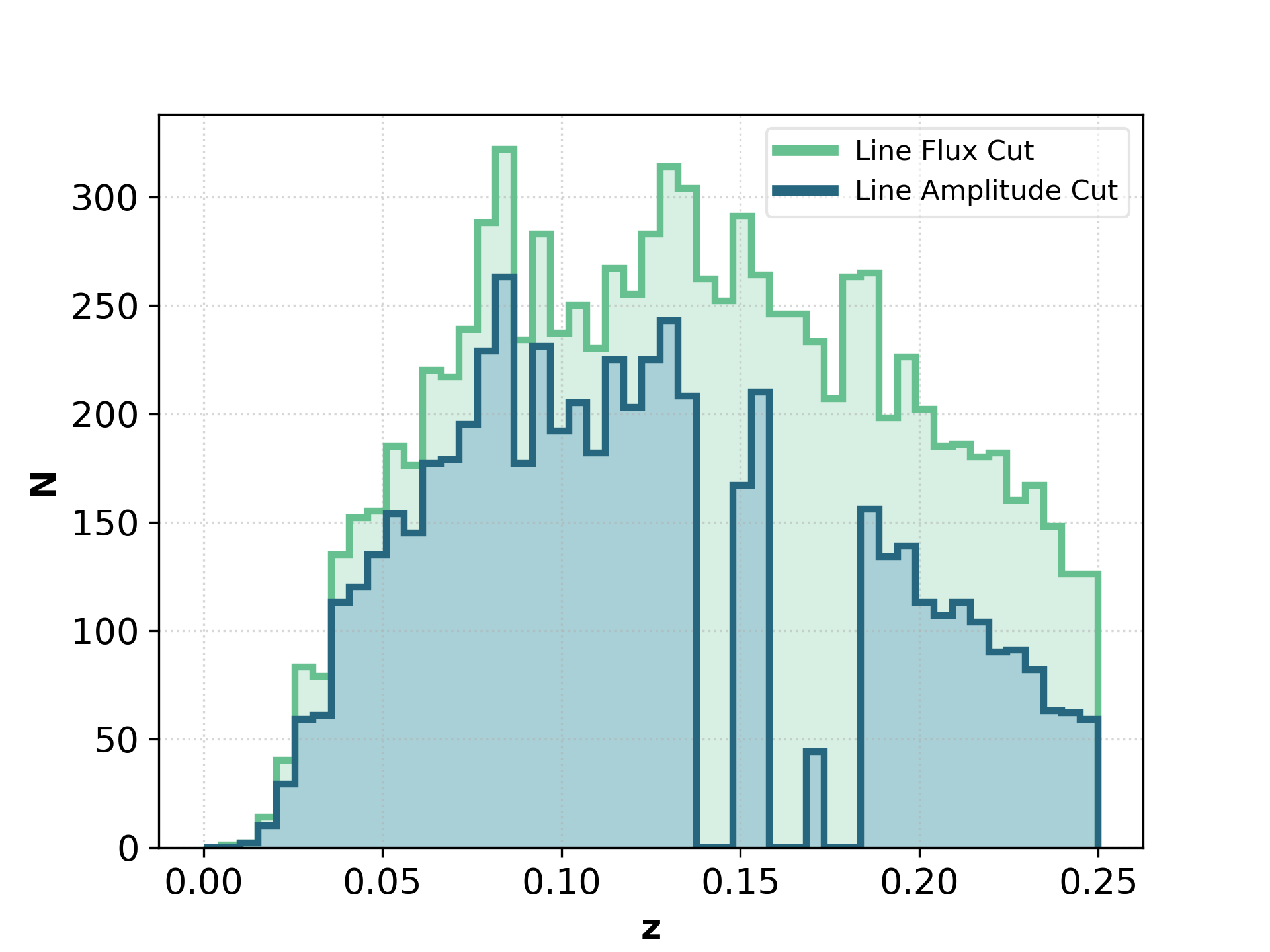} 
\caption{The redshift distribution of our parent galaxy sample, shown once with the signal to noise cut performed on line fluxes and again with the cut performed on line amplitudes. The line flux uncertainties in DESI are systematically underestimated due to an incomplete propagation of said uncertainty. This effect is particularly apparent for lines falling in the DESI spectral overlap region, which for H$\alpha$ and H$\beta$ occur at 0.14 $<$ z $<$ 0.18. As such, we use the more restrictive SNR cut on the line amplitudes. All cuts described in Table 1 are applied in this plot; the two samples differ only in how the SNR cut was performed. The sample produced using cuts on line flux contains 9580 galaxies; the sample produced using cuts on line amplitudes contains 5696 galaxies.}
\label{fig:redshift-histogram}
\end{figure}

We note that the Balmer line fluxes provided by \texttt{IRON} are corrected for stellar absorption. The package used to derived the line fluxes (and other galaxy parameters; e.g. stellar mass and star formation rate) in \texttt{IRON}, \texttt{fastspecfit}, used a combination of FSPS \citep{FSPS} stellar continuum spectra to fit to the combined spectrophotometric DESI data. Although this correction for stellar absorption is rudimentary, it is still sufficient as even if it produces a systematic offset in $\tau_{b}$, the attenuation curve depends on the \textit{difference} in Balmer optical depths and so our output curves will not be significantly affected \citep{Battisti2016}. As described below, we control for stellar population age via a cut on $D_n4000$.

Because we look at metallicity-related trends, and because we want to exclude galaxies containing AGNs (based on the BPT diagram), we also require that our galaxies have a SNR$ > 3$ in the [N II] $6584$ $\mathrm{\AA}$  line as well as the [O III] $5007$ $\mathrm{\AA}$  line. Like the Balmer lines, this signal-to-noise cut is performed on the line amplitudes. The subsequent BPT cut is performed on the \citet{kaufBPT} line for purely star-forming galaxies. An additional SNR$ > 3$ cut is performed on the [O II] $3726$ line, but only when that line is used to calculate the ionization parameter for the curves in Section 4.3.3.

\begin{deluxetable*}{c|c}
\tablecaption{Sample Selection Summary} 
\tablehead{Parameter & Value}
\startdata
    \multicolumn{2}{c}{Cuts From DESI Catalogs} \\
    \hline
    Line SNR (H$\alpha$, H$\beta$) & $>$ 3.00 \\
    Line SNR ([N II]$6584$, [O III]$5007$) & $>$ 3.00 \\
    \texttt{ZWARN}& 0\\ 
    \texttt{SPECTYPE} & \texttt{GALAXY}\\
    \texttt{MORPHTYPE} & none of \texttt{PSF}, \texttt{DUP}, \texttt{DEV}\\ 
    D$_n$4000 & 1.1 - 1.3\\
    $\tau_b$ & 0.00 - 0.95 \\
    BPT Diagram & \citet{kaufBPT}, star-forming branch \\
    $f_{\mathrm{enclosed}}$ & $>$ 0.33 \\
    z & $<$ 0.25 \\
    \hline
    \multicolumn{2}{c}{Cuts From GALEX Catalogs} \\
    \hline
    SNR (FUV) & $>$ 3\\
    SNR (NUV) & $>$ 3\\
    \texttt{FOV\_RADIUS} & $<$ $0\fdg55$\\
    \texttt{FUV\_ARTIFACT} & none of \texttt{4, 32, 64}\\
    \texttt{NUV\_ARTIFACT} & none of \texttt{2, 4, 32, 64}\\
\enddata  
\label{tab:schecParams}
\tablecomments{Line SNRs are calculated on amplitudes. $f_{\mathrm{enclosed}}$ is the fraction of luminosity from each galaxy enclosed by the DESI fiber.  }
\end{deluxetable*}

We also make use of the built-in \texttt{IRON} quality and classification flags. We require that all sources have \texttt{ZWARN=0} (corresponding to sources associated with no known artifacts) and \texttt{SPECTYPE=GALAXY} (which removes spectra that look like stars or other non-galactic sources, and additionally acts as a secondary way to remove AGNs). After matching the \texttt{IRON} catalog to the \texttt{DESI\_DR1.PHOTOMETRY} catalog from NOIRLAB DATALAB (corresponding to data from the Legacy Surveys DR9) \citep{legsurvey}, we apply additional cuts based on the \texttt{TRACTOR}-derived galaxy morphologies. We reject galaxies where \texttt{morphtype=PSF}, \texttt{morphtype=DUP}, or \texttt{morphtype=DEV}. The first selection removes sources with a PSF-like morphology (i.e. stars, if not already removed by the \texttt{SPECTYPE} cut). The second forces the catalog to use the morphology measured by TRACTOR in the case that a source is associated with a GAIA measurement. The third cut removes sources that are best-fit with a De Vaucouleurs profile. This is appropriate because galaxies that are best-fit with such a profile are typically elliptical and not actively star-forming, and we wish to construct our sample out of star-forming galaxies. 

We perform an additional cut based on the fraction of light from each galaxy enclosed by the DESI fiber. There is a significant size mismatch between the (small, $\sim 1.5^{\prime\prime}$ diameter) DESI fiber and the (large, $\sim 4.5^{\prime\prime}$ FWHM) GALEX PSFs. This makes matching the two problematic for very extended galaxies, as DESI and GALEX will be measuring fundamentally different regions of these galaxies which may contaminate the attenuation signal. To mitigate this issue, we calculate the fractional flux enclosed by the DESI fiber based on the light profile fits in \texttt{DESI\_DR1.PHOTOMETRY} and reject any galaxies where this fraction is less than 0.33. This selection is made at 0.33 instead of a higher fraction like 0.5 as a compromise to retain a usable sample size.

Finally, we reject any galaxies with a $D_n4000$ less than 1.1 or greater than 1.3. This range encloses the peak of our $D_n4000$ distribution while being narrow enough to produce reliable curves (Figure 2) \citep{Battisti2016}. Specifically, this restricts us to systems with a light-weighted stellar population age ranging approximately from 0.1 to 1 Gyr \citep{Kauffmann2003,Chen2009}. One of the key assumptions underlying the pairwise method of making attenuation curves is that the effective unattenuated spectrum of each galaxy is (nearly) identical. Though in practice this cannot be guaranteed, it is still best practice to use galaxies of comparable age (and thus intrinsic spectral shape). Indeed, previous work has established that attempting to derive attenuation curves without enforcing such a cut on spectral shape will produce curves with artificially high attenuation in the UV as a result of the intrinsically redder nature of older galaxies \citep{Battisti2016}. Our tests of more restrictive $D_n4000$ cuts, e.g., limiting to the range 1.1-1.2 or 1.2-1.3, did not improve or significantly change the results, but only increased scatter in the derived attenuation curves. Additionally, galaxies with Balmer optical depth $<0.1$ are not used in the construction of our attenuation curves.

\begin{figure}[htb!]
\centering
\includegraphics[width=1.0\columnwidth]{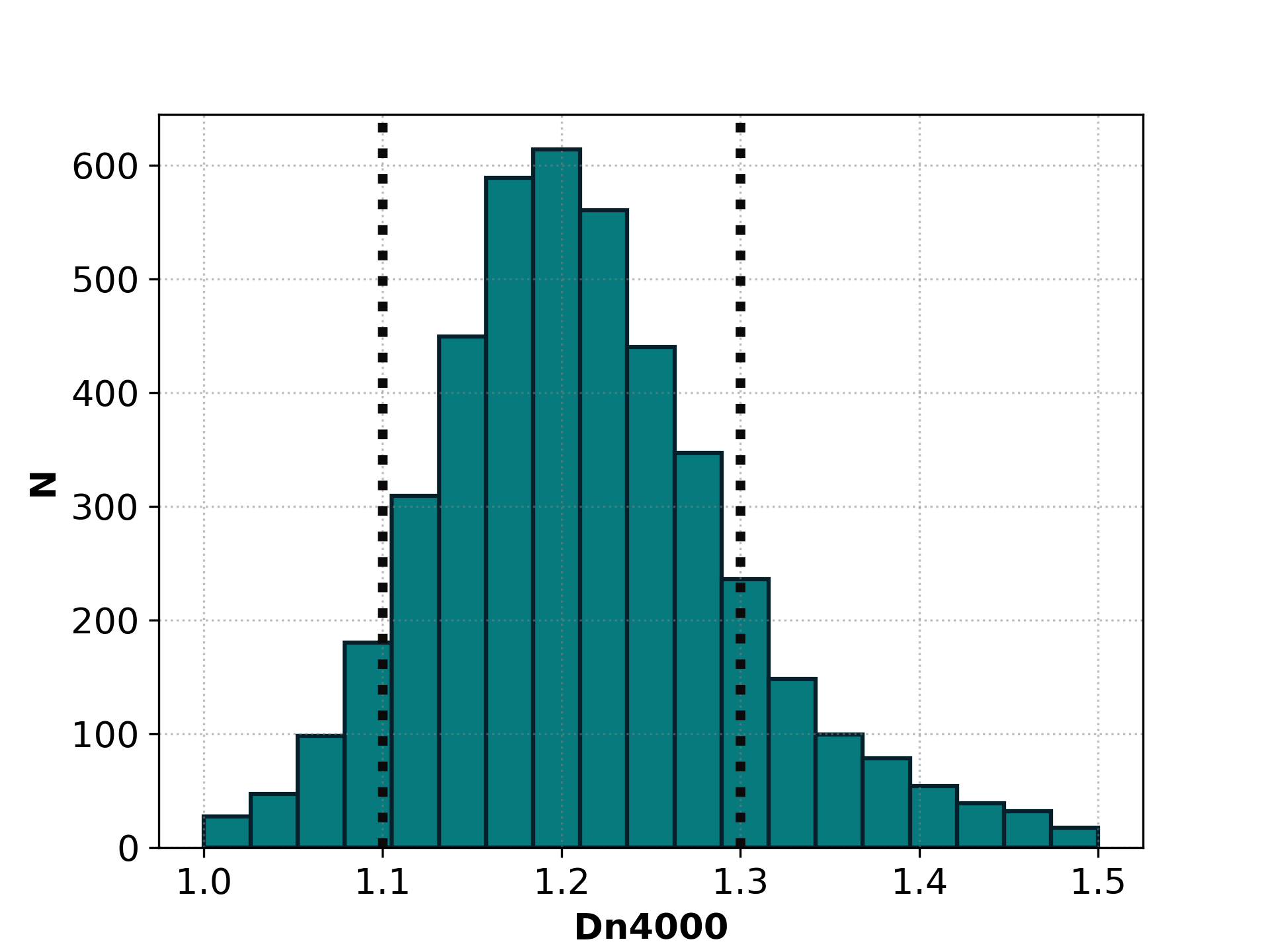} 
\caption{The $D_n$4000 distribution of our sample once all other cuts have been applied. The peak of this distribution lies between 1.1 and 1.3, which are the boundaries of the cut used in this work.}
\label{fig:redshift-histogram}
\end{figure}

\subsubsection{GALEX Quality Cuts}
To ensure only quality GALEX measurements were selected, we follow the general approach used by \citet{Bianchi17}, but applied to the GALEX MIS (in \texttt{GUVCAT\_MCS}) sources rather than the GALEX \texttt{AIS} sources. The \texttt{AIS} catalogs have been found to be biased towards very blue galaxies, an issue which the deeper MIS catalog avoids \citep{Battisti2016}. We apply an SNR $>$ 3 cut on both the NUV and FUV. We further require that all sources be within $0\fdg55$ of the center of the instrument FOV (which has radius $0\fdg60$), as rim artifacts become common as one approaches the edge of the detector; the quality of astrometry and photometry tends to decrease near the edge as well. 

Furthermore, we reject any sources with \texttt{FUV\_artifact} = 4, 32 or 64, \texttt{NUV\_artifact} = 4, 32 or 64 and / or \texttt{NUV\_artifact} = 2. Targets with \texttt{artifact}=4 or 64 suffer from dichroic reflection, while targets with \texttt{artifact}=2 suffer from window reflection. \texttt{artifact}=32 corresponds to rim artifacts. The FUV detector is not susceptible to window reflection artifacts. Per \citet{Bianchi17} and the GALEX DR6 documentation, these are the only artifact flags of significant concern. Finally, we reject any remaining duplicates after these cuts are applied by selecting the GALEX source with centroid closest to the DESI fiber center, if multiple GALEX targets were associated with a single DESI target. This is uncommon -- even papers that matched GALEX photometry to fiber spectra within a much larger radius (5$^{\prime\prime}$) found that duplications were rare \citep{Salim16}. Throughout this paper, we use the GALEX forced photometry at a $4.5^{\prime\prime}$ diameter.

\subsubsection{Matching and Sample Size}
To be considered a match, we require that the distance between the center of a DESI target and the center of a GALEX target is less than the DESI fiber diameter (1.50$^{\prime\prime}$). The GALEX quality cuts described above were performed before cross-matching with DESI; after matching and pruning duplicates as described, there were 92,937 candidate galaxies. This is reduced to 65,718 after SNR cuts on the Balmer lines, 31,962 after SNR cuts on [N II]$6584$ and [O III]$5007$, and 26,994 after applying the BPT cut. The cuts on \texttt{SPECTYPE, ZWARN} and morphology reduced this to 26,079. This step decreases the sample minimally from the previous set of cuts because the selection of SNR of the emission lines already removes the vast majority of elliptical galaxies. The redshift cut further reduced this to 24,600 sources. The cut on $f_{enclosed}$ reduced this to 7494 sources, and the $D_n4000$ cut reduced it further to 5696 galaxies. Of these, 3457 have $\tau_b$ $>$ 0.1 and were used in the construction of our attenuation curves.

\section{Methods}
We produce our selective attenuation curves ($k_\lambda - R_V$) using the method originally presented in \citet{Calzetti94} and later expanded upon by \citet[][among others]{reddy15,Battisti2016,shivaei2020}. To summarize, we begin by selecting galaxies from our parent sample based on physical criteria of interest (e.g. stellar mass, metallicity, sSFR). Each of these samples is itself divided into bins as a function of Balmer optical depth $\tau_{b}$. Then, we produce median stacks of the galaxies within each of these $\tau_b$ bins. Attenuation curves are then derived from the difference between stacks at different $\tau_b$. We describe a necessary pre-processing step, aperture correction, in Section 3.1 below. Section 3.2 describes the strategy we use to bin and stack our galaxies as a function of $\tau_b$. Finally, Section 3.3 describes how we use these stacks to derive our attenuation curves.

\subsection{DESI-GALEX Aperture Correction}
The DESI fiber (1.5$^{\prime\prime}$ diameter) and the GALEX PSFs (4.9$^{\prime\prime}$ FWHM in the NUV, 4.2$^{\prime\prime}$ FWHM in the FUV) are not the same size. Observations performed using these instruments will fundamentally provide information on different physical scales, no matter what corrections are applied. That said, applying an approximate aperture correction is still justified. When generating our stacked spectra (as described in Section 3.3), all data for each galaxy are normalized relative to their rest-frame $5500\AA$ flux density -- so, to produce realistic stacks, the correction used to match GALEX photometry to the DESI spectroscopy has to be done consistently. To do this, we correct both the DESI fiber fluxes and GALEX photometry to an effective 4.5$^{\prime\prime}$ diameter aperture. This correction is also required in order to accurately measure the UV slope, which can be biased due to the difference in size between the 2 GALEX filter PSFs.

\begin{figure*}[htb!]
  \centering
  \includegraphics[width=0.49\textwidth]{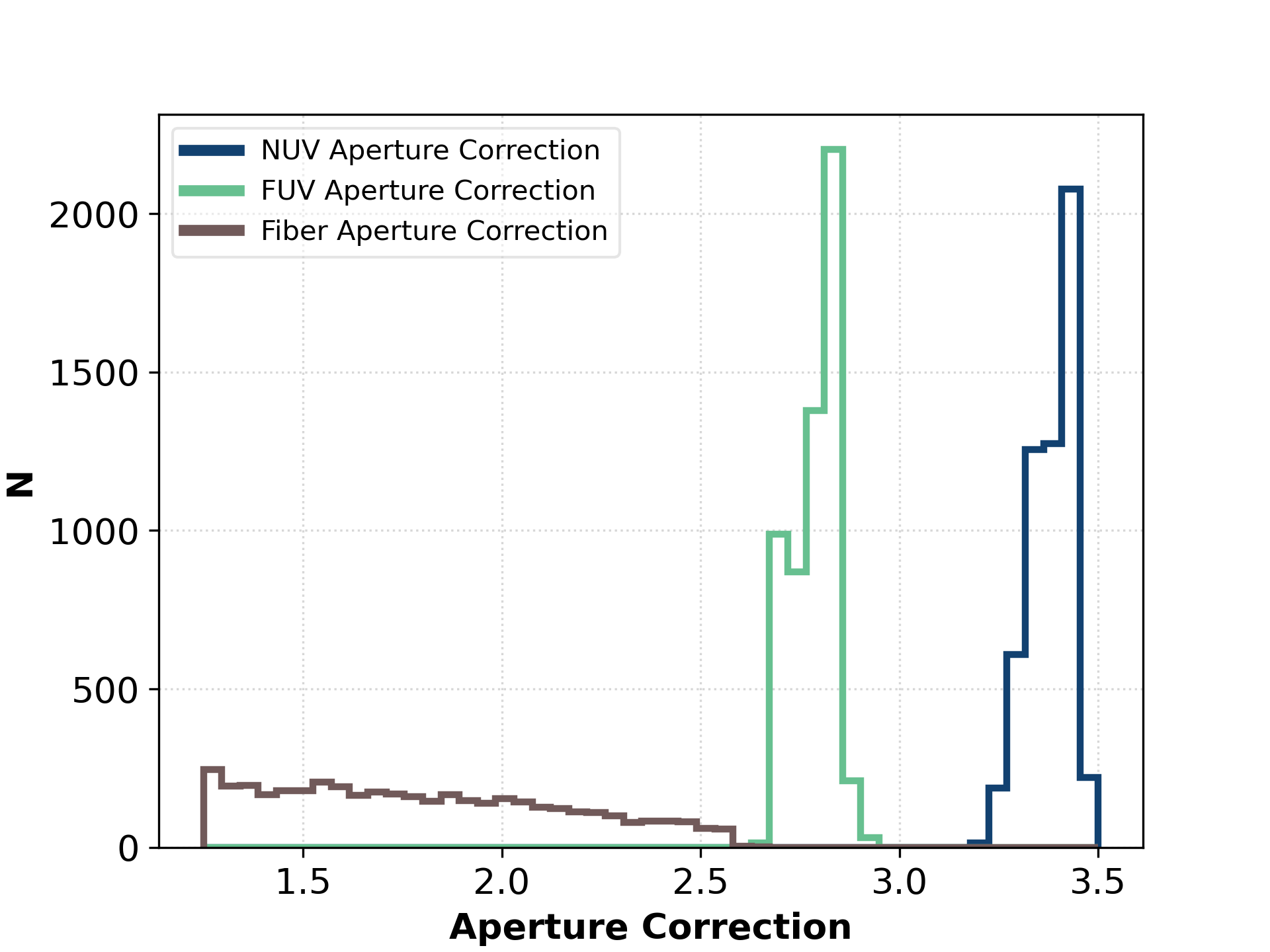}\hfill
  \includegraphics[width=0.49\textwidth]{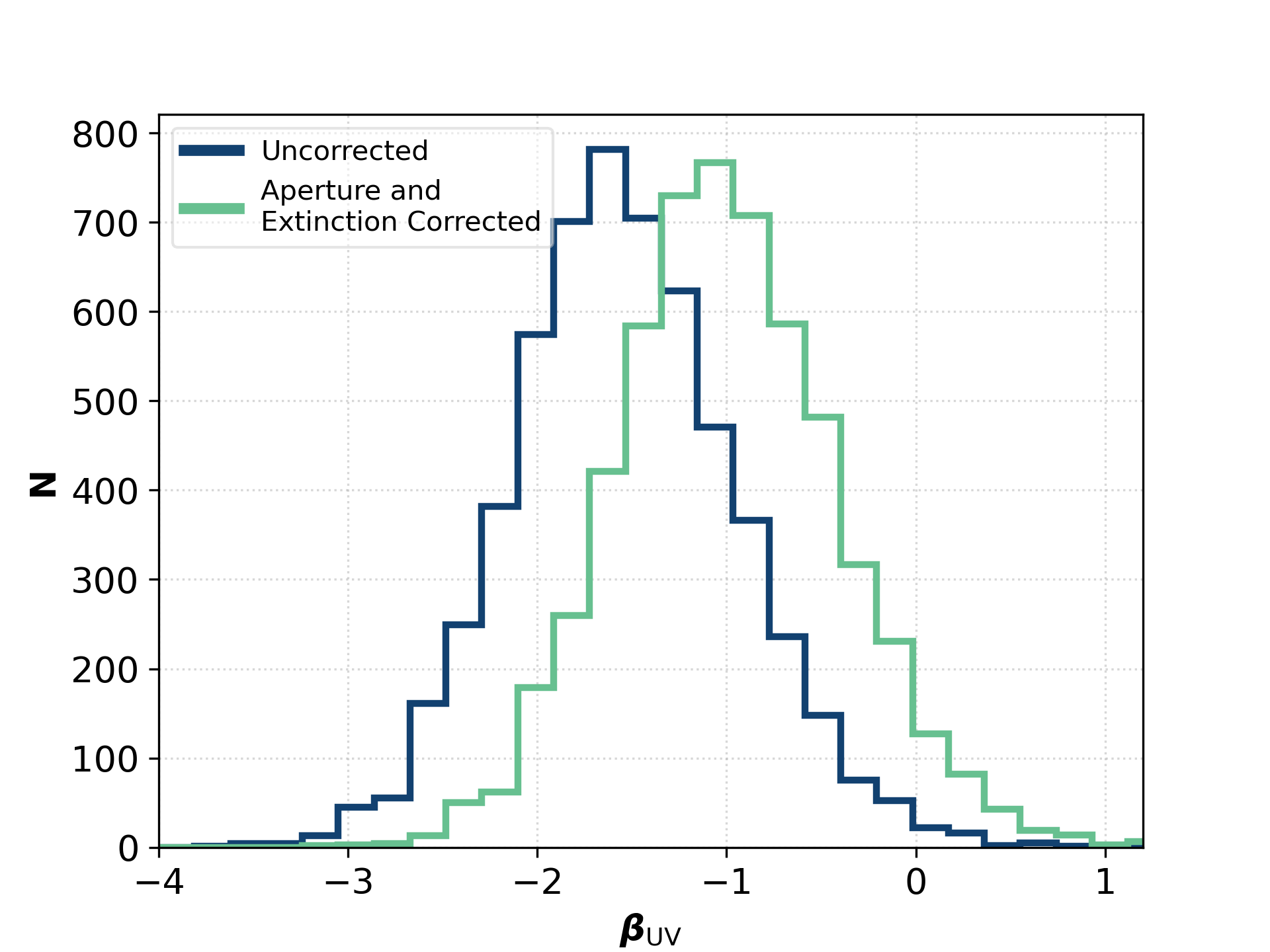}
  \caption{(a) Histograms of the aperture corrections we derived for the DESI fiber, GALEX FUV filter, and GALEX NUV filter. In all cases, the data are corrected to a 4.5$^{\prime\prime}$ diameter effective aperture. (b) Histograms of the observed UV slopes, before and after aperture and extinction corrections were applied. Because the GALEX has a broader NUV PSF than it does in the FUV, applying such a correction reddened the average UV slope.}
  \label{fig:apcorr}
\end{figure*}

After the morphological cuts performed in Section 2.2.1, we are left with galaxies corresponding to one of three MORPHTYPEs: SER (a general Sersic profile), REX (a "round exponential" profile; i.e. a Sersic profile with n=1 and an ellipticity of 0), and EXP (an n=1 exponential profile with nonzero ellipticity) \citep{legsurvey}. We note that the Legacy Surveys perform these morphological fits on a combination of g, r, and z-band imaging. A single profile based on a combination of these 3 bands is provided for each galaxy. We reconstruct these 2D intensity profiles on a grid of $50\times50$ pixels, with a pixel scale of $0.5^{\prime\prime}$.  

We then derive the appropriate corrections in each band based on these intensity profiles. For DESI, this is straightforward: the correction is 
{\abovedisplayskip=6pt \belowdisplayskip=6pt \abovedisplayshortskip=4pt \belowdisplayshortskip=4pt
\begin{equation}
\mathcal{C}_{\rm FIBER}
= \frac{F(R_{\rm ap})}{F(R_{\rm fib})}.
\end{equation}}
Here, \(F(R_{ap})\) is the flux from the intensity profile within a 4.5$^{\prime\prime}$ diameter aperture, while \(F(R_{fib})\) is the same but within a 1.5$^{\prime\prime}$ aperture. This is analogous to how \citet{Battisti2016} performed their fiber aperture correction.
Deriving the GALEX aperture corrections requires one additional step. The GALEX PSFs have very broad wings that cannot be approximated by a Gaussian. We thus produce a model of each source as it would be seen in each GALEX filter by taking the original model of each source and convolving it with the empirically derived GALEX NUV or FUV PSFs\footnote{\url{https://www.galex.caltech.edu/researcher/techdoc-ch5.html}} as appropriate. The aperture correction is then
{\abovedisplayskip=6pt \belowdisplayskip=6pt \abovedisplayshortskip=4pt \belowdisplayshortskip=4pt
\begin{equation}
\mathcal{C}_{\rm GALEX}
= \frac{F(R_{\rm ap})}{F_{\rm FUV \vee NUV}(R_{\rm ap})}.
\end{equation}}
where, once again, \(R_{\rm ap}=4.5\arcsec\). Histograms of the fiber, FUV, and NUV aperture corrections are provided in Figure 3a. The GALEX NUV PSF is broader than the GALEX FUV PSF. As a consequence, the NUV data require a larger aperture correction than the FUV (Figure 3a) and the resulting (corrected) UV slope becomes redder (Figure 3b). 

We note that because the DESI fiber has a fixed size, we are unable to account for radial variation in $\tau_b$. It is well established that galaxies show gradients in dust content such that their centers tend to be more heavily obscured \citep{Nelson16,Greener20}. Thus, the Balmer decrements derived only apply to the regions of our galaxies enclosed within the fiber and may not be perfectly representative of each galaxy as a whole. This cannot be accounted for with an aperture correction and should be kept in mind as a limitation of our work. It should also be kept in mind that assuming an optical light profile for the UV emission, as we do here, represents another significant limitation as light profiles may be attenuation-dependent.

\subsection{Binning and Stacking Spectra}
One of the challenges of this work was the fact that performing a cut on stellar mass, metallicity, or ionization parameter also represents a secondary cut on $\tau_b$. To mitigate this effect, we do not use the same bin edges in $\tau_b$ for each curve we made, but instead create evenly spaced $\tau_b$ bins across each distribution, while ensuring each bin contains a reasonable number of galaxies. This is a novel approach.

To reiterate, we do not use the same $\tau_b$ bins for each curve we produce from the stacked spectra; rather, we assign bin edges based on the $\tau_b$ distribution of the galaxies that would be used in that curve. As some galaxy properties are correlated with dust content through $\tau_b$ using identical $\tau_b$ bins for different sub-samples would either result in some bins containing an insufficient number of galaxies to produce usable stacks, or have required us to throw out a large number of galaxies. 

In sources with $\tau_b \leq 0$, $\tau_b$ does not contain any useful information about the attenuation in a given galaxy and so should not be used to derive attenuation curves. We also need to reject galaxies that are optically thick in the UV or in the nebular emission lines. To conservatively enforce both of these criteria, we reject galaxies with $\tau_b < 0.1$ or $\tau_b > 0.95$. It has been found that galaxies with $\tau_b < 0.1$ (which we might naively associate with little to no obscuration) have an average UV continuum attenuation of $A_{FUV} = 1.3$ \citep{Salim18}. We found that including galaxies with $\tau_b < 0.1$ led us to produce attenuation curves that were subject to significantly increased scatter. 

After performing this cut, the galaxies in each sample are separated into 4 bins of $\tau_b$. Every sample from which an attenuation curve is constructed was partitioned into these bins by first calculating the 95$\%$ quantile of its $\tau_b$ distribution, then separating the resulting $\tau_b$ range into 4 bins of equal size between $\tau_b$ = 0.1 and the $\tau_b$ of the 95$\%$ quantile. This allowed us to produce reasonably populated bins of $\tau_b$ for each sub-sample without throwing away a large number of galaxies. Stacked spectra are produced after binning by $\tau_b$.


When generating curves from galaxy samples that have been subject to cuts on some parameter (e.g. metallicity) it is sometimes enforced that the resulting $\tau_b$ distributions are matched (between sub-samples), as it is suspected that the shape of the attenuation curve may vary as a function of total attenuation \citep[e.g.][]{shivaei2020}. This allows one to claim that any observed variation in curve shape is due to the probed parameter alone, rather than due to a secondary correlation with opacity. Although well-justified, we elect not to perform such cuts in this work. This enables us to answer similar questions from the opposite direction: are there parameters that correlate strongly with opacity that are \textit{not} correlated with the shape of the attenuation curve?

\begin{figure}
  \centering
  \includegraphics[width=\columnwidth]{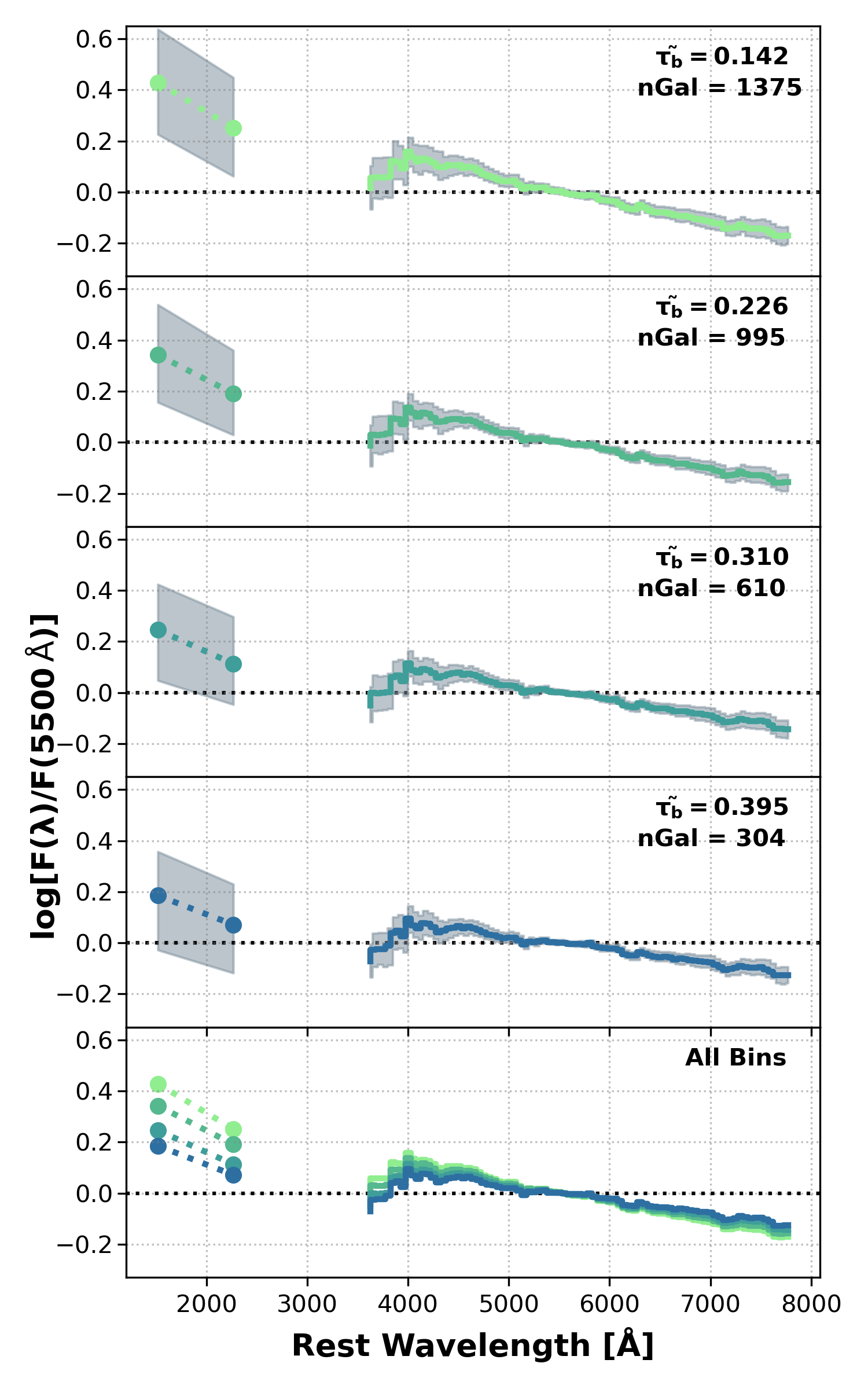}
  \caption{The stacked spectra used to produce the attenuation curve shown in Figure 7.  The y-axis in each panel is $\log[F(\lambda)/F(5500\,\mathrm{\AA})]$; shaded bands mark intervals enclosing 68\% of the population used to make each stack. There is significantly more scatter associated with the UV photometry than there is associated with the spectra, perhaps because of the positional uncertainty inherent to matching the large GALEX PSF to the DESI fiber. Despite the underlying scatter, the UV points match up sufficiently well with the optical spectra that we are satisfied with our aperture corrections.}
  \label{fig:stacks_all}
\end{figure}

After binning, we produce a median stacked spectrum from the galaxies within each bin. Individual spectra were retrieved using NOIRLab's SPectra Analysis and Retrievable Catalog Lab (\texttt{SPARCL}) client \citep{SPARCL}. We then iterate through the sample on a bin-by-bin, spectrum-by-spectrum basis. First, the GALEX NUV and FUV photometry are attached to each spectrum and the previously-discussed aperture corrections are applied. Emission lines in each spectrum are removed by replacing a $40\mathrm{\AA}$ region around each line with realistic noise sampled from a (sigma-clipped) $500\mathrm{\AA}$ region centered on that line. The spectrum and its associated UV photometry are then normalized to the value of the spectrum at $5500\mathrm{\AA}$ (i.e. V band). To better resolve the continuum shape, we bin the spectra in 50$\mathrm{\AA}$ intervals (excluding the UV photometry).

Once each individual spectrum is processed, we calculate the median spectrum in each bin. We use a 1000-run Monte Carlo resampling to estimate the uncertainty on this median stack. We also calculate the $68\%$ enclosed region around each point in the spectrum and UV photometry in order to get a sense of the scatter associated with these measurements. A set of example stacks are shown in Figure 4. We find that though the scatter can be significant, particularly for the UV photometry and for points in the spectrum near the 4000 angstrom break, the stack medians are quite stable. 

As our sample spans the redshift range 0.0$\sim$0.25, the DESI fiber subtends between $<$1~kpc and $\sim$6~kpc depending on the distance of the galaxy. We checked the impact of redshift on our derivation by dividing our sample in 2 redshift bins and comparing the results. The differences between derived curves were within our uncertainties, indicating that we can treat the entire 0.0$\sim$0.25 redshift range as a single distance bin. Details of this test are shown in Appendix C.

\subsection{Making Attenuation Curves}
Given any two distinct templates, $F_j(\lambda)$ and $F_k(\lambda)$, associated with different values of $\tau_b$, the selective attenuation curve is \begin{equation} Q_{j,k} = \frac{-\ln\frac{F_j(\lambda)}{F_k(\lambda)}}{\tilde\tau_{b,j} - \tilde\tau_{b,k}} \end{equation} where $\tilde\tau_{b,j}$ and $\tilde\tau_{b,k}$ are the median Balmer optical depths of the galaxies in stack $j$ and stack $k$ respectively. We calculate $Q_{j,k}$ for each unique pair of stacks. We estimate uncertainties in these selective curves by performing another (1000-run) Monte Carlo resampling. The effective attenuation curve $Q_{eff}$ is then the median of all $Q_{j,k}$ for a given sample. 

$Q_{eff}$ only contains information about the difference in attenuation between any two wavelengths. It is related to the more useful total attenuation curve $k_\lambda$ via \begin{equation} k_\lambda = fQ_{eff}+R_V \end{equation} Here, $f$ accounts for the difference between nebular and stellar reddening. This correction is made by altering the slope of the attenuation curve such that $k_\lambda(B) - k_\lambda(V) = 1$: \begin{equation} f = \frac{1}{Q_{eff}(B) - Q_{eff}(V)} \end{equation}  It can also be expressed in terms of the differential reddening between the nebular and stellar emission: \begin{equation} f = E(B-V)_{nebular}/E(B-V)_{stellar} \end{equation} Our curves are constructed such that by definition, $Q_{eff}(V) = 0$. To calculate the uncertainty on f for the median curve, we first determine the value of f$_{j,k}$ for each $Q_{j,k}$. The reported uncertainty on f is then the standard deviation of the set of all f$_{j,k}$.

$R_V$ is the total-to-selective attenuation; in the formalism we use, it can be thought of the total curve's vertical offset from 0 at $5500\mathrm{\AA}$. It is typically calculated either based on an energy-balance argument \citep[e.g.][]{calzetti}, or by measuring the curve in the near-infrared \citep[][]{shivaei2020,Battisti17}. In this work, we derive $R_V$ via the extrapolation method after fitting to the underlying attenuation curves. 

We model $k_\lambda$ as a single polynomial in 
$1/\lambda$. In most cases, a 3rd order polynomial is sufficient to produce a reasonable fit. However, some of the curves we produce are better represented by a 2nd order polynomial, most likely due to scatter in the UV. Our general model is thus \begin{equation} k_\lambda = (c_3 x_{\mu m^{-1}}^3 + c_2 x_{\mu m^{-1}}^2 + c_1 x_{\mu m^{-1}} + c_0) + R_V \end{equation} where $ x_{\mu m^{-1}} = 1/\mu m$. When a 2nd order polynomial is used, $c_3$ is set to zero. 

\section{Results}
We report our results below and in Table 2. We describe the properties of our sample in Section 4.1. The attenuation curve produced using the entirety of our sample is presented in Section 4.2. Section 4.3 focuses on how -- or whether -- the shape of the attenuation curve depends on various galaxy properties.

\subsection{Properties of The Sample}
\begin{figure}[t!]
\centering
\includegraphics[width=1.0\columnwidth]{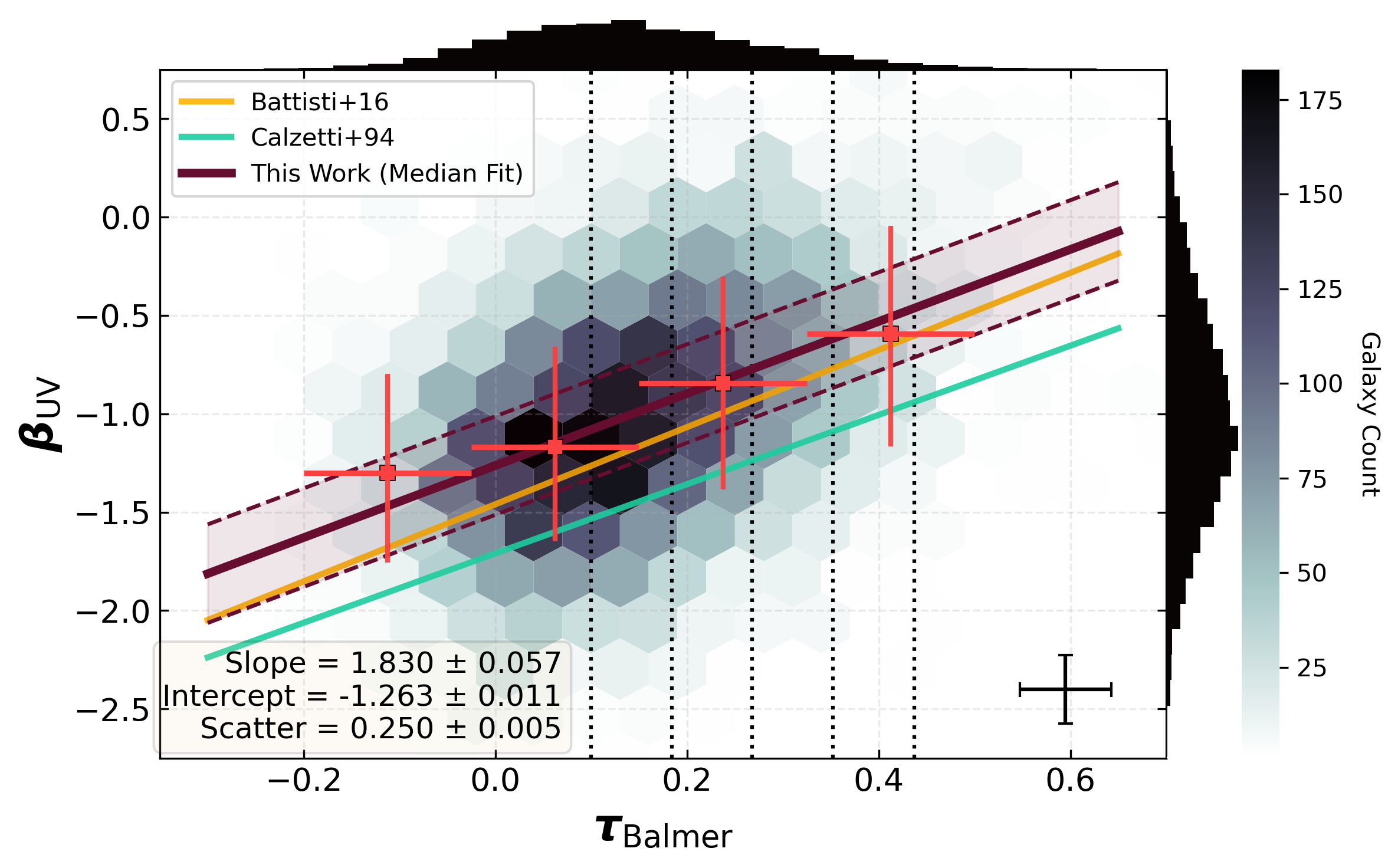} 
\caption{UV continuum slope versus Balmer optical depth for our galaxies. The maroon line and shaded region represents our galaxies (best fit and intrinsic scatter); the turquoise is the best-fit line from \citep{Calzetti94}, and the orange line is the best-fit from \citep{Battisti2016}. The points are the binned median values of $\beta_{UV}$ as a function of $\tau_b$ with the error bars corresponding to the $68\%$ enclosed region in the $\beta_{UV}$ direction, and the bin width in the $\tau_b$ direction. The vertical dotted lines show the bins used in curve creation. Representative median errorbars are shown in the lower right. 1D distributions of both variables are shown along the edges of the plot. Though we find a similar relationship between $\beta_{UV}$ and $\tau_b$ as other work, we find a slightly weaker correlation driven perhaps by our UV and optical data probing significantly different physical scales.}
\label{fig:balmer-v-beta}
\end{figure}
The pairwise method of deriving attenuation curves stemmed originally from the observation that the UV stellar continuum reddening ($\beta_{UV}$) is correlated to the reddening of nebular emission lines ($\tau_b$) \citep{Calzetti94}. In Figure 5, we show the relationship between these two parameters in our data. 

\begin{figure*}[htb!]
  \centering
  \gridline{
    \fig{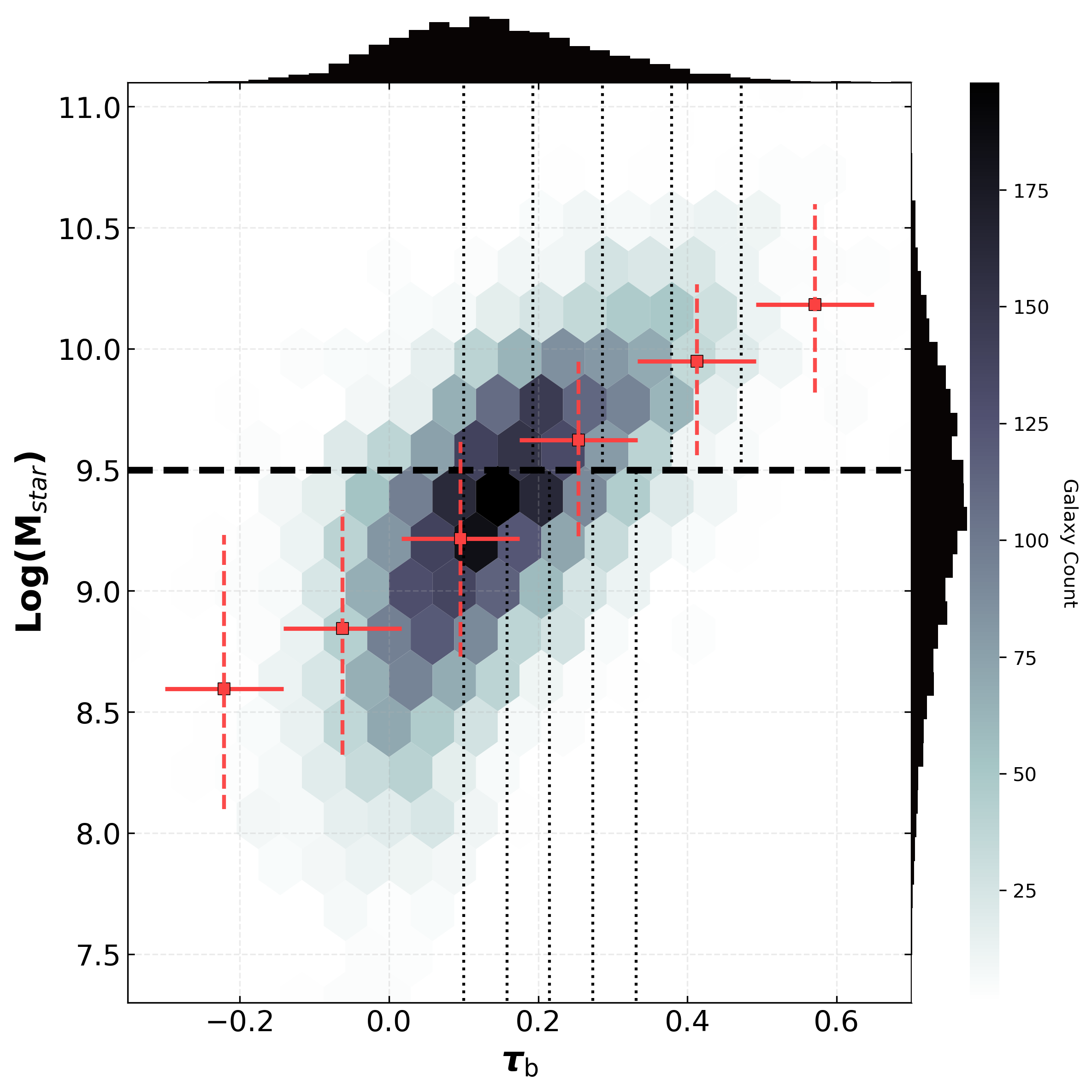}{0.48\textwidth}{(a) $\tau_b$ vs. Mass}
    \fig{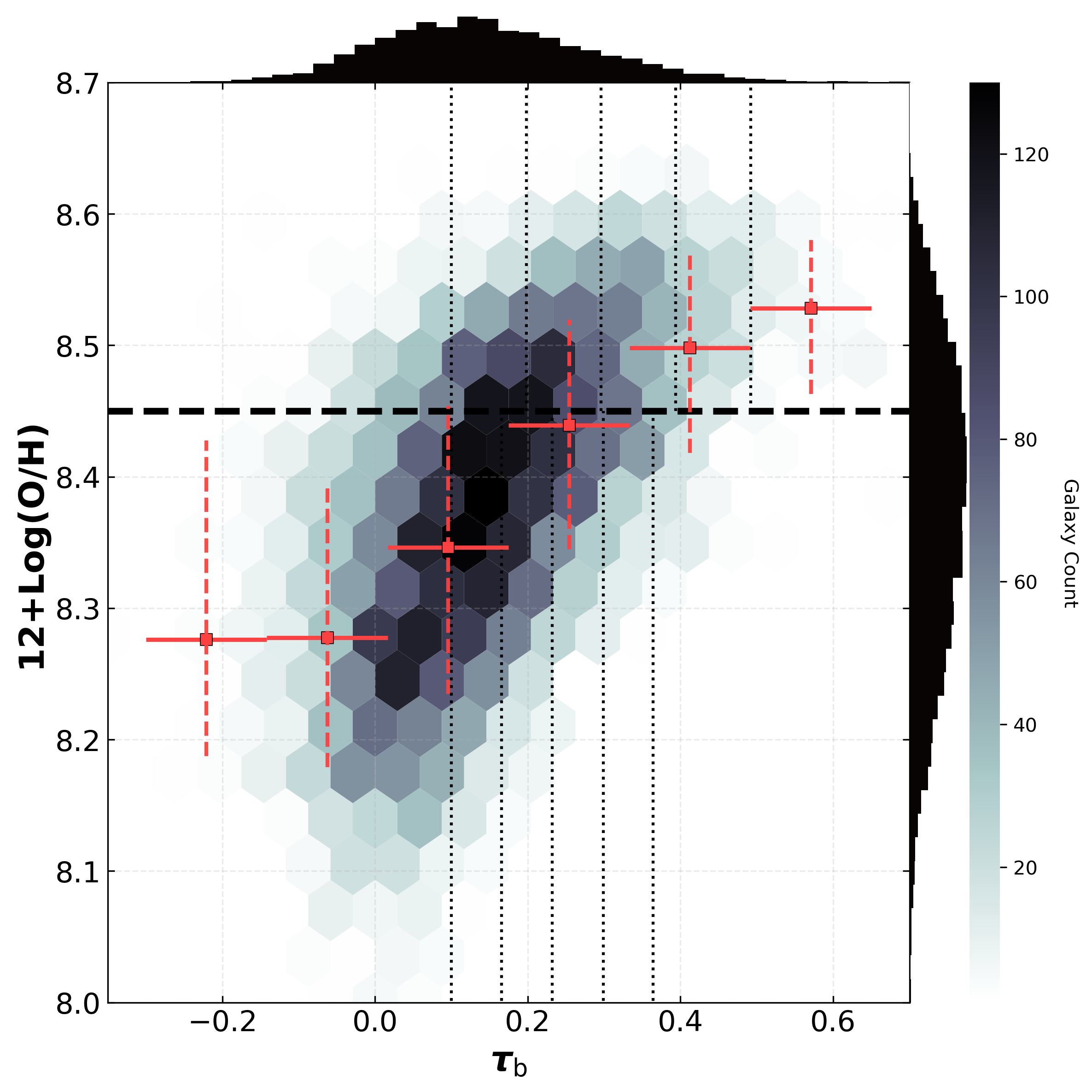}{0.48\textwidth}{(b) $\tau_b$ vs. Metallicity}
  }
  \gridline{
    \fig{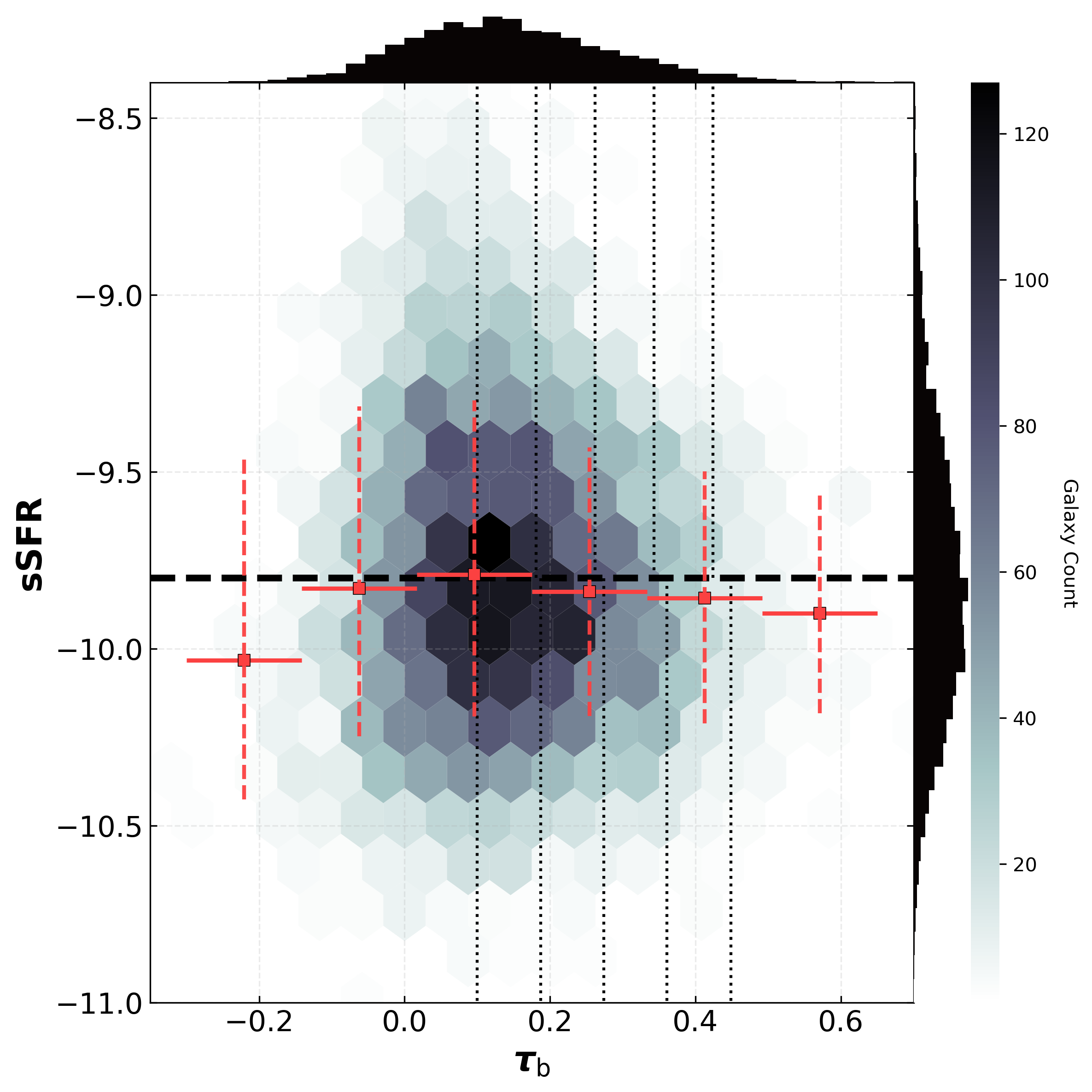}{0.48\textwidth}{(c) $\tau_b$ vs. sSFR}
    \fig{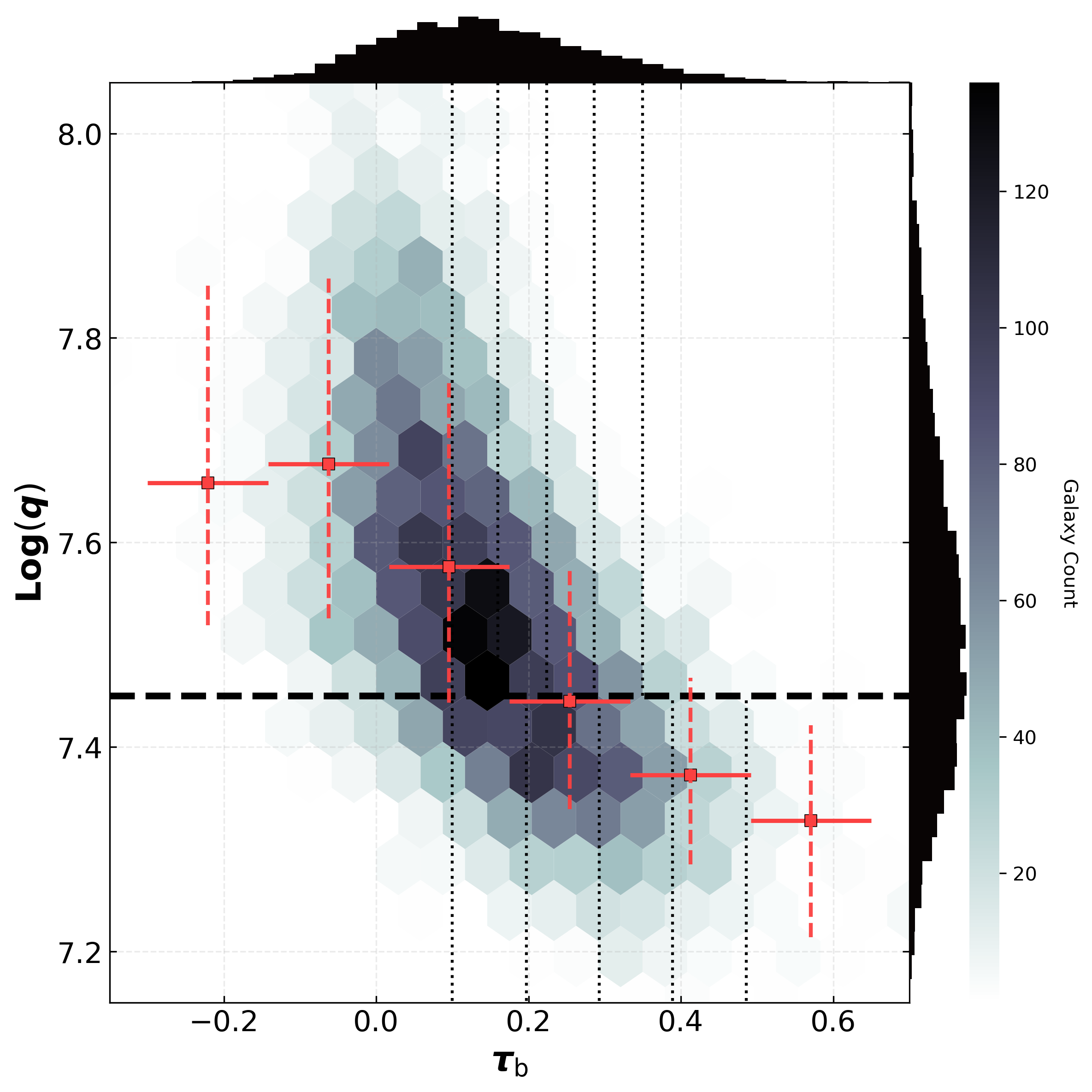}{0.48\textwidth}{(d) $\tau_b$ vs. Ionization Parameter}
  }

  \caption{$\tau_b$ vs. the four galaxy parameters examined in this study: stellar mass (a), metallicity (b), sSFR (c) and ionization parameter (d). We find a strong correlation between $\tau_b$ and the parameter of interest for mass, metallicity, and ionization parameter. In contrast, sSFR does not appear to be strongly correlated with $\tau_b$. Although galaxies with negative $\tau_b$ are not used to construct our attenuation curves, we show them here for completeness. The red points show the rolling median of each distribution; the vertical dashed errorbars represent the $68\%$ enclosed region and the horizontal solid errorbars show the range of $\tau_b$ covered by each point. The horizontal dashed lines show where each distribution was split into two sub-samples; the vertical lines denote the $\tau_b$ bins used to create the relevant attenuation curve.}
  \label{fig:fourpanel}
\end{figure*}

We find a relationship between $\tau_b$ and $\beta_{UV}$ consistent with previous work. The functional behavior of this relationship depends on the star/dust geometry; typically, this is found to be linear. We fit it using the Python package \texttt{linmix} \citep{linmix}, which uses a hierarchical Bayesian approach to perform linear regression on data with uncertainty in both $x$ and $y$. This gives \begin{equation}\beta_{UV} = (1.83 \pm 0.06)\tau_b - (1.26\pm 0.01)\end{equation} with intrinsic scatter $\sigma_{int} = 0.25 \pm 0.01$. A Spearman R test returns a $p$-value of 0.39, indicating a (somewhat weak) positive correlation. The scatter in this relation may in part be due to the difference in angular scales probed between DESI and GALEX; no matter what aperture correction is applied, the data represent emission phenomena arising on different scales. 

We also examine the relationship between $\tau_b$ and other galaxy properties -- specifically metallicity ($12+\mathrm{Log(O/H)}$), the ionization parameter (log(q)), stellar mass ($\log(M_{\star})$), and specific star formation rate (sSFR). We calculate metallicities using the \citet{o3n2} O3N2 calibration. Ionization parameters are based on the \citet{KK04} parameterization of the \citet{KD02} relation. We use the catalog stellar masses. Star formation rates are derived from the \citet{Kennicutt98} $H\alpha$ indicator, adjusted to use a \citet{Chab} IMF. We note that the absolute accuracy of any of these measurements is not important to the results of this work, as we are only interested in the properties of our galaxies relative to one another. For instance, we verified that using a differently calibrated metallicity indicator (such as the O3N2 or N2 calibrations derived in \citet{PP04}) might produce a different average metallicity but does not affect the rank-ordering of galaxies by metallicity. We find that the metallicity tracer used does not significantly impact the relative behavior of our attenuation curves (Appendix B). 

In Figure 6, we plot the four galaxy properties described above -- stellar mass, metallicity, sSFR, and ionization parameter -- against $\tau_b$. Our sample spans a relative wide range in each of these parameters -- approximately 7.5 - 10.5 in $\log(M_{\star})$, 8.0-8.6 in $12+\mathrm{Log(O/H)}$, 7.2-8.0 in $\mathrm{Log(q)}$, and -11.0 to -8.5 in sSFR. We find that 3 out of 4 (stellar mass, metallicity, and ionization parameter) are correlated with $\tau_b$, with Pearson R values of 0.65, 0.59, and -0.63 respectively. The specific star formation rate, meanwhile, is not correlated with $\tau_b$, with a Pearson R of -0.04. This is because both star formation rate and stellar mass correlate with $\tau_b$, which cancels out the trend in sSFR. We note that the sSFRs we report are fiber sSFRs -- when calculating sSFR, the stellar mass is corrected for the fraction of light within the fiber, based on the Sersic fits. Whether or not we do this does not impact the interpretation of our results, however. We show the value of each parameter at each sample was split as the horizontal dashed line, and the bin edges used to make curves as the vertical dashed lines. When making curves split by a parameter correlated with $\tau_b$, we find that the most reliable curves are made when the split is slightly offset from the center in order to allow the low-$\tau_b$ subsample to span a wider range of $\tau_b$ than would otherwise be possible.

\begin{deluxetable}{lrrrrrrrrr}
\tablecaption{Coefficients for $k_\lambda$}
\tablehead{
\colhead{Curve} &
\colhead{$c_3$} & \colhead{$c_2$} & \colhead{$c_1$} &
\colhead{$c_0$} & \colhead{$f$} & \colhead{$R_V$} &
\colhead{$N_{gal}$}
}
\startdata
Bulk Curve & $0.02\pm0.01$ & $-0.41\pm0.10$ & $3.67\pm0.26$ & $-5.41\pm0.23$ & $2.52\pm0.15$ & $3.97\pm0.27$ & 3284 \\
\tableline
Log(O/H)+12 $>$ 8.45 & $0.06\pm0.02$ & $-0.86\pm0.19$ & $4.97\pm0.52$ & $-6.56\pm0.42$ & $3.73\pm1.16$ & $4.76\pm1.52$ & 1360 \\
Log(O/H)+12 $<$ 8.45 & $0.04\pm0.02$ & $-0.65\pm0.19$ & $4.33\pm0.57$ & $-5.97\pm0.48$ & $3.21\pm0.30$ & $4.36\pm0.51$ & 1923 \\
\tableline
Log(q) $>$ 7.45 & $0.05\pm0.03$ & $-0.80\pm0.26$ & $4.80\pm0.76$ & $-6.40\pm0.61$ & $3.02\pm0.66$ & $4.63\pm1.10$ & 1740 \\
Log(q) $<$ 7.45 & $0.05\pm0.02$ & $-0.81\pm0.20$ & $4.90\pm0.53$ & $-6.51\pm0.42$ & $3.64\pm1.20$ & $4.77\pm1.60$ & 1539\\
\tableline
Log($M_\star$) $>$ 9.5 & $-----$ & $-0.26\pm0.01$ & $3.27\pm0.08$ & $-5.07\pm0.10$ & $2.57\pm0.22$ & $3.75\pm0.34$ & 1791\\
Log($M_\star$) $<$ 9.5 & $-----$ & $-0.28\pm0.02$ & $3.33\pm0.13$ & $-5.14\pm0.16$ & $2.09\pm0.27$ & $3.66\pm0.50$ & 1492\\
\tableline
sSFR $>$ -9.8 & $-----$ & $-0.18\pm0.02$ & $2.95\pm0.10$ & $-4.76\pm0.07$ & $2.45\pm0.36$ & $3.52\pm0.53$ & 1539\\
sSFR $<$ -9.8 & $0.04\pm0.01$ & $-0.65\pm0.15$ & $4.40\pm0.41$ & $-6.08\pm0.33$ & $2.72\pm0.52$ & $4.44\pm0.88$ & 1744 \\
\enddata
\tablecomments{For each curve, ($c_3$, $c_2$, $c_1$) and ($c_0$) are the fit coefficients. $c_3$ is set to zero if the curve is better fit by a 2nd order polynomial. $f$ accounts for the difference in reddening between the nebular emission and stellar continuum. The uncertainty reported for $f$ is the standard deviation of the set of f values derived for all individual $Q_{n,r}$.} $R_V$ is the extrapolated total-to-select extinction. $N_{gal}$ is the number of galaxies used in each curve.
\label{tab:coeffs_labeled_bulk}
\end{deluxetable}

\subsection{The Bulk Attenuation Curve}

\begin{figure}[htb!]
\centering
\includegraphics[width=1.0\columnwidth]{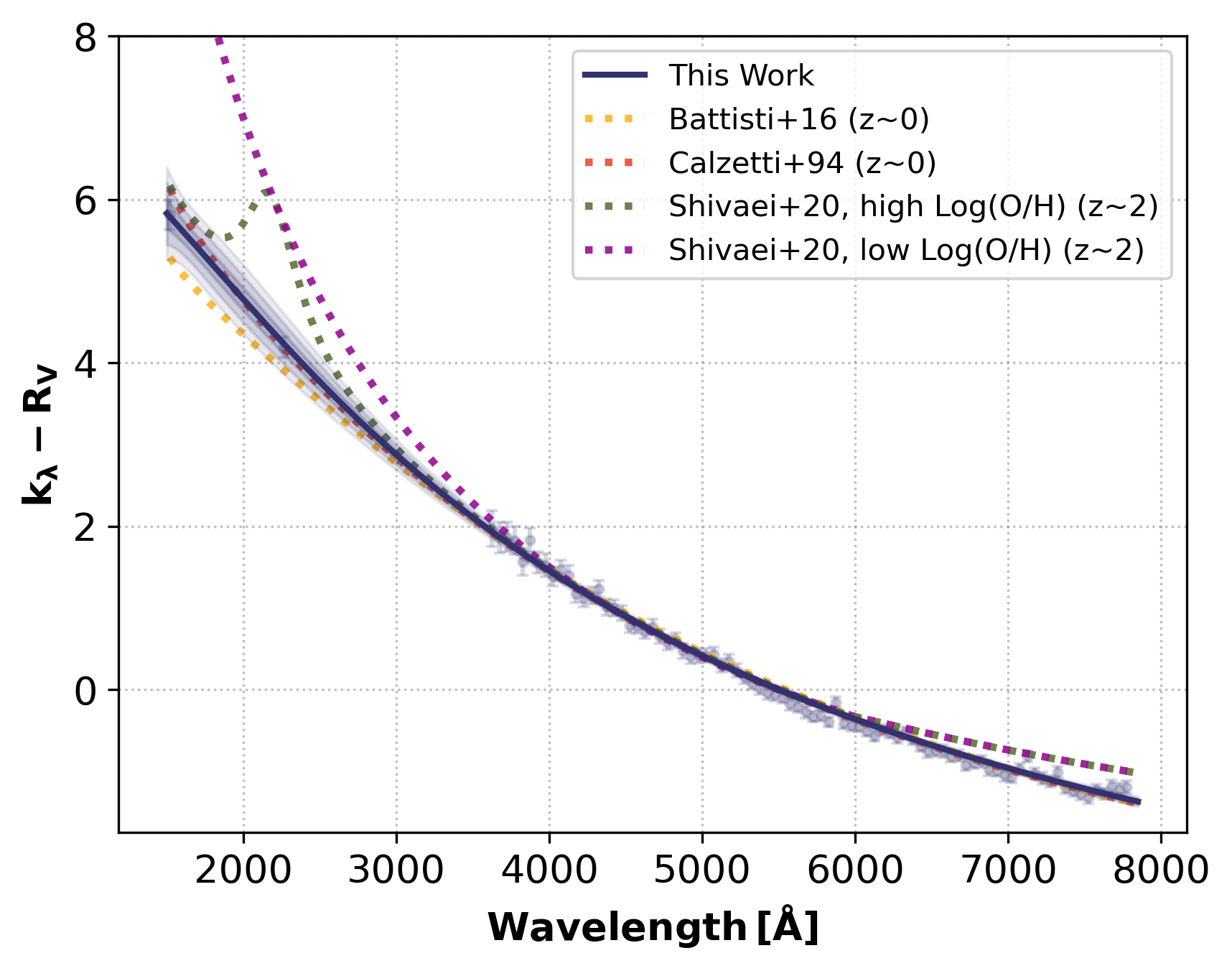} 
\caption{Comparison of our selective attenuation curve (dark blue line) against a number of other curves from the literature. The dotted yellow line corresponds to \citet{Battisti2016}, the dotted red line corresponds to \citet{Calzetti94}, and the dotted green line corresponds to the high-metallicity subsample from \citet{shivaei2020}. Our selective attenuation curve is steeper in the UV than the \citet{Battisti2016} curve and is most similar to the curve of \citet{Calzetti94}. Regions of 1, 2, and 3$\sigma$ uncertainty in our fit are shown as the shaded areas around the curve. The points are the underlying empirical attenuation curve from which the fit is produced.}
\label{fig:tlbEvo}
\end{figure}

We first present our `bulk' attenuation curve -- i.e, the curve produced using all galaxies in our sample with no further cuts applied. We note that throughout this paper, when plotting our attenuation curves, we show normalized selective attenuation curves (i.e. $k_\lambda - R_V$) rather than total attenuation curves. This allows us to emphasize differences in curve shapes and elides the (typically somewhat uncertain given the lack of IR coverage in our data) $R_V$, which we still provide for the sake of completeness (in Table 2). We find a bulk attenuation curve that is extremely similar to the results of \citet{Calzetti94}. The coefficients used to describe this curve is given in Table 2. The curve itself is shown alongside a set of representative curves from the literature in Figure 7. 

We find a curve that is well-described by a smooth 3rd-order polynomial and which has a shape consistent with the attenuation curve derived in \citep{Calzetti94,calzetti}. The curve has $R_V = 3.97 \pm 0.27$, consistent with the value found by \citep{calzetti} ($4.05\pm0.80$), and $f = 2.52 \pm0.27$. This value of $f$ is consistent with other studies performed in the low-redshift Universe \citep{calzetti,Battisti2016}. 

 
\subsection{How Does The Attenuation Curve Vary With Galaxy Properties?}
The next 4 subsections focus on whether the curve varies as a function of metallicity, stellar mass, the ionization parameter, and the specific star formation rate. To do this, we split our sample in two at the median value of each parameter and produce an attenuation curve using each sub-sample. We note that in each case, the relative behavior of the split curves is not sensitive to where upper/lower split is placed.

\begin{figure}[htb!]
\centering
\includegraphics[width=1.0\columnwidth]{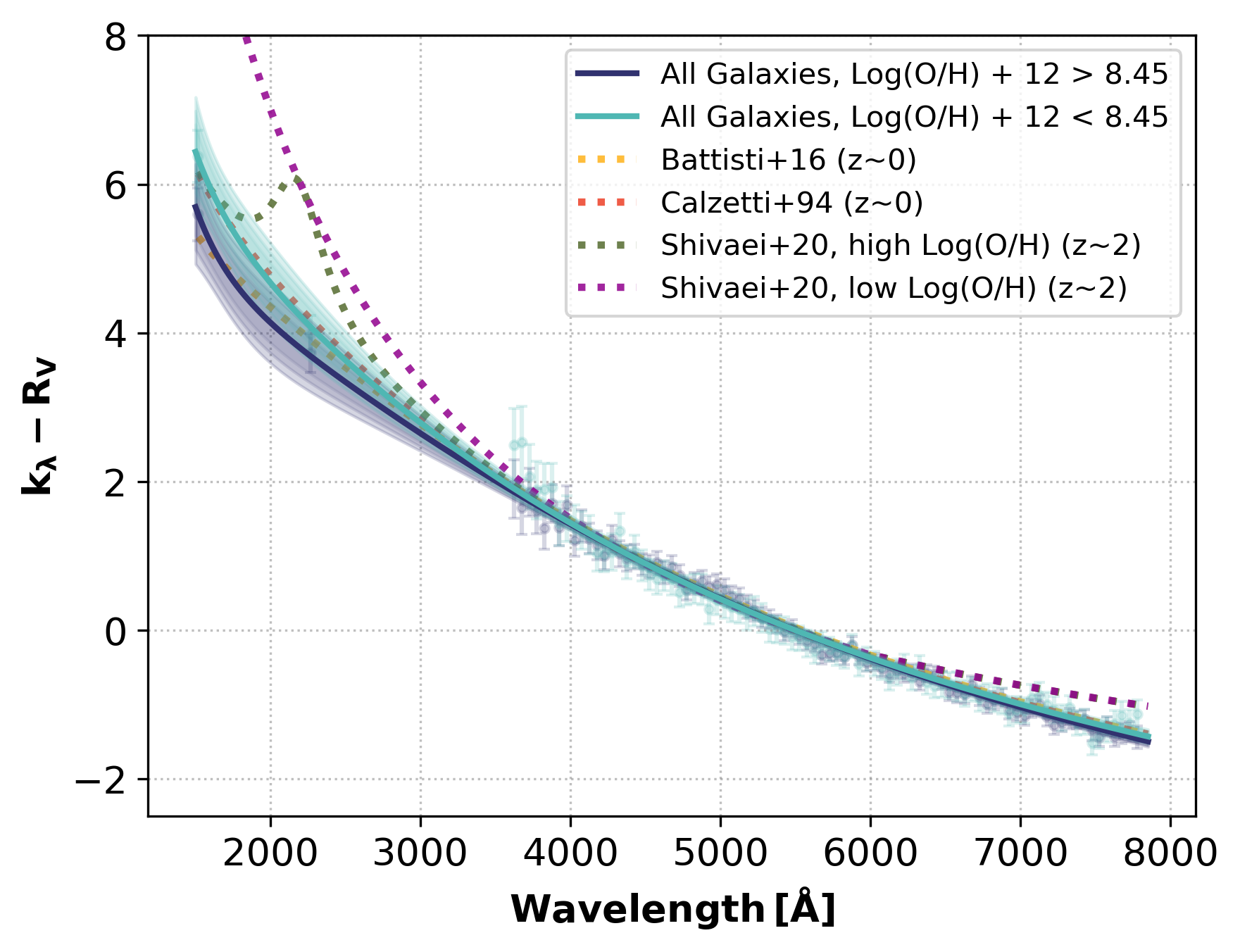} 
\caption{Metallicity-binned attenuation curves split at Log(O/H) + 12 = 8.45. We find no significant difference in shape between our low-metallicity and high-metallicity attenuation curves.}
\label{fig:metals}
\end{figure}
 
\subsubsection{Metallicity}
The first galaxy property we examine is metallicity. Metallicity is relatively well studied in an attenuation context; if variation in attenuation curve shapes is due primarily to differences in grain properties, rather than differences in average star-dust geometry, then metallicity might be expected to correlate strongly with curve shape. Recent work suggests that the curve steepens as one pushes to systems with lower metallicity, at least at high redshift \citep{shivaei2020}. This slope difference is observed to occur even after controlling for opacity. The picture is less clear in the local universe, where some work has found no link between gas-phase metallicity and the attenuation curve slope \citep{Salim18}. To test this in our data, we split our galaxy sample at Log(O/H)+12 = 8.45, and construct attenuation curves as before for each of the resulting two bins. We show these curves in Figure 8 and their parameters in Table 2.

Unlike \citet{shivaei2020}, we do not find a significant difference between our low-metallicity and high-metallicity curves. Our low-metallicity curve does appear to be slightly steeper, but if real this effect is very mild. We also find a different relationship between metallicity and nebular-to-stellar reddening than \citet{shivaei2020}; they found that their low-metallicity sample was subject to about twice as much differential reddening as their high-metallicity sample, while we find a similar amount of differential reddening in both curves -- possibly slightly more in the high metallicity case. The values of $f$ we find for both metallicity bins are slightly higher than our bulk curve, but are relatively uncertain. This propagates through to the values of $R_V$, which are also slightly elevated but uncertain enough that they remain consistent with \citep{calzetti}. We note, however, that the sample of \citet{shivaei2020} spans a wider metallicity range than our sample does ($12+\mathrm{Log}(O/H) = 8.0 - 8.8$). It was also constructed using a different metallicity diagnostic (the NII/H$\alpha$ diagnostic of \citet{PP04}, which tends to produce systematically higher metallicities) making a one-to-one comparison difficult. It is possible that the lack of evolution we see is due to a sample that does not span a sufficient metallicity range. 

\subsubsection{Stellar Mass}

\begin{figure}[htb!]
\centering
\includegraphics[width=1.0\columnwidth]{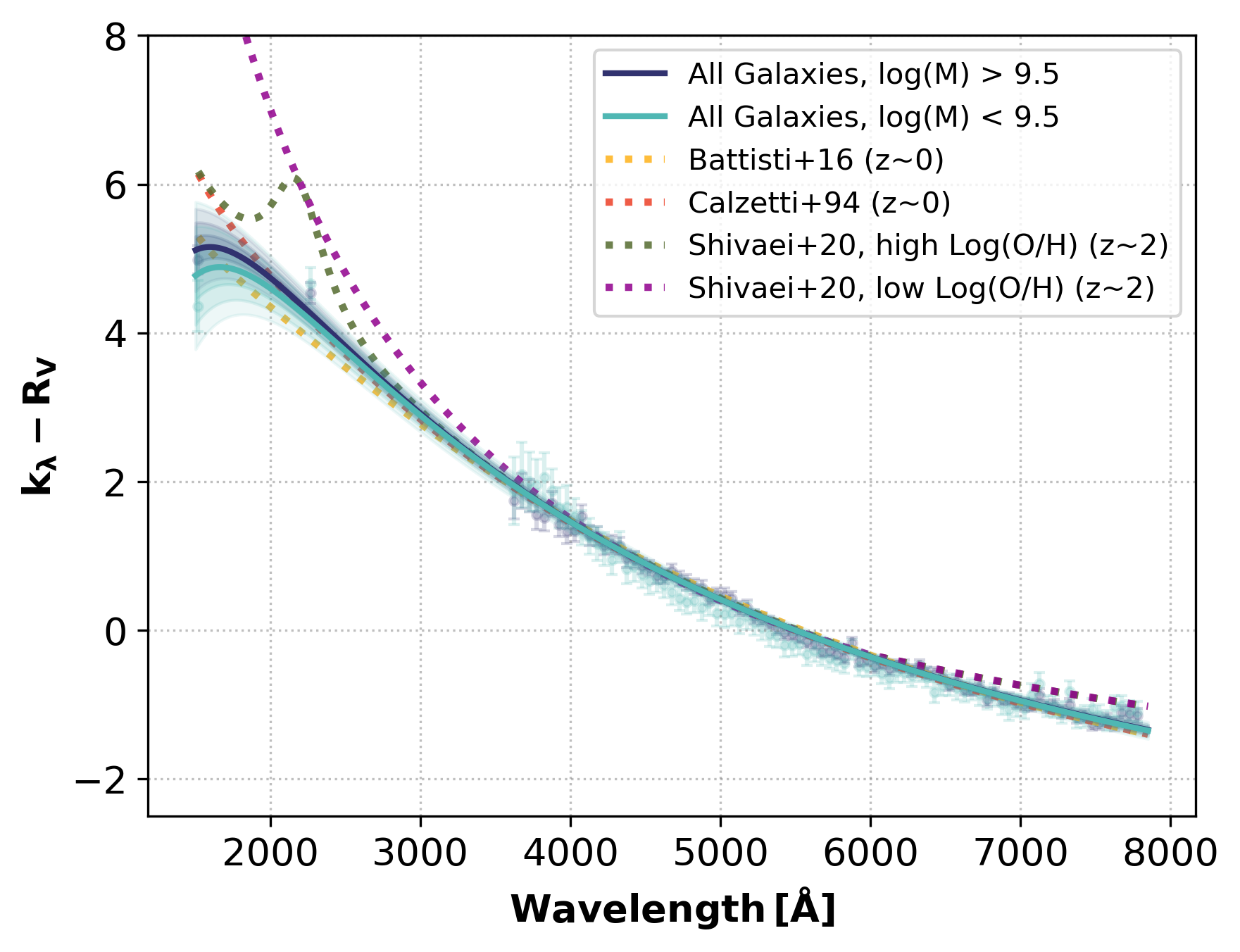} 
\caption{Mass-binned attenuation curves split at Log($M_\star$) = 9.5. Once again, we find no significant difference between the shape of these curves. The somewhat 'flat' shape of both curves in the UV is likely driven by scatter in either the FUV, NUV, or both.}
\label{fig:mass}
\end{figure}

The shape of the attenuation curve might also correlate with galaxy stellar mass, with more massive galaxies corresponding to shallower curves. There is evidence of this at redshifts $>1$; the observed UV slope reddens significantly between Log($M_\star$) $\sim$8.0 and $\sim$11.5 \citep{Battisti22}. Such a trend is also visible in some studies at low redshift. This is probably a consequence of the mass-metallicity relationship. More massive star-forming galaxies tend to be more metal rich (as well as more rich in gas, a vital ingredient for the presence of dust), and these more metal-rich galaxies tend to be more dust-rich.

To test this in our data, we again split our sample in two: this time at its median stellar mass of Log[$M_\star$] = 9.5. The resulting curves are shown in Figure 9; the relevant fit parameters are in Table 2. We find that there is no significant relationship between stellar mass and curve slope. The two curves have values of $f$ and $R_V$ that are consistent both with one another and the \citet{calzetti} values. Both of the mass-binned curves we produce are slightly 'flat' in the UV (the FUV is slightly suppressed compared to the NUV), but this is of low significance given the uncertainty on these points and so is likely driven by scatter.

\subsubsection{Ionization Parameter}
\begin{figure}[htb!]
\centering
\includegraphics[width=1.0\columnwidth]{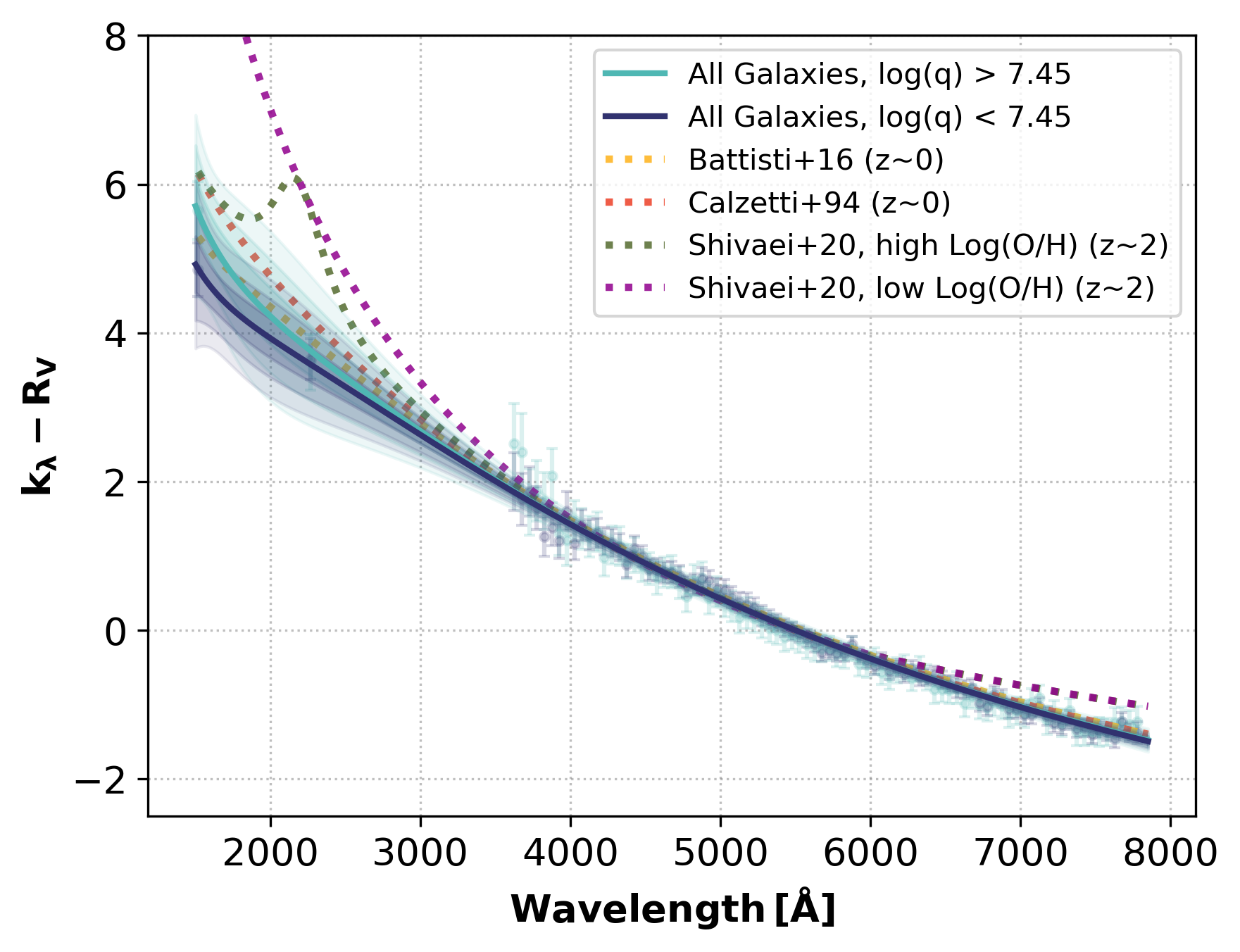} 
\caption{Log(q)-binned attenuation curves split at Log(q) = 7.45. As with stellar mass and metallicity, we find no strong link between the ionization parameter and the shape of the attenuation curve.}
\label{fig:lq}
\end{figure}
As a tracer of the intensity of the UV radiation field, the ionization parameter may provide insight into dust grain processing. The presence of a strong radiation field may destroy grains through photodesorbption \citep{Draine1979} or rotational disruption \citep{Hoang2019}; there is also the possibility of inhibited grain growth \citep{Whitcomb2024,Whitcomb2025}. Although UV radiation is probably not the dominant mechanism destroying grains in most galaxies (sputtering in gas-grain collisions likely dominates \citep{Jones2011}), it might be significant enough that galaxies hosting different average UV radiation fields might host different average dust grain size distributions. Small grains are better at absorbing short-wavelength photons, so with all else being held equal one may expect steeper attenuation curves in environments rich in small grains. However, as the ionization parameter is strongly correlated with metallicity, it may be difficult to distinguish between effects arising from the two.  In modeling-based, high-redshift studies, the ionization parameter has been found to be only weakly correlated with the attenuation curve shape \citep{Markov25}.

To test this, we split our sample at Log(q) = 7.45. The resulting curves are shown in Figure 10; their parameters are given in Table 2. Once again, we find no significant difference between the curve shapes. The curves have values of $f$ and $R_V$ that are consistent between themselves as well as with the \citep{calzetti} curve. In our data, it does not appear the the ionization parameter is strongly correlated with the shape of the attenuation curve. 

\subsubsection{sSFR}
\begin{figure}[htb!]
\centering
\includegraphics[width=1.0\columnwidth]{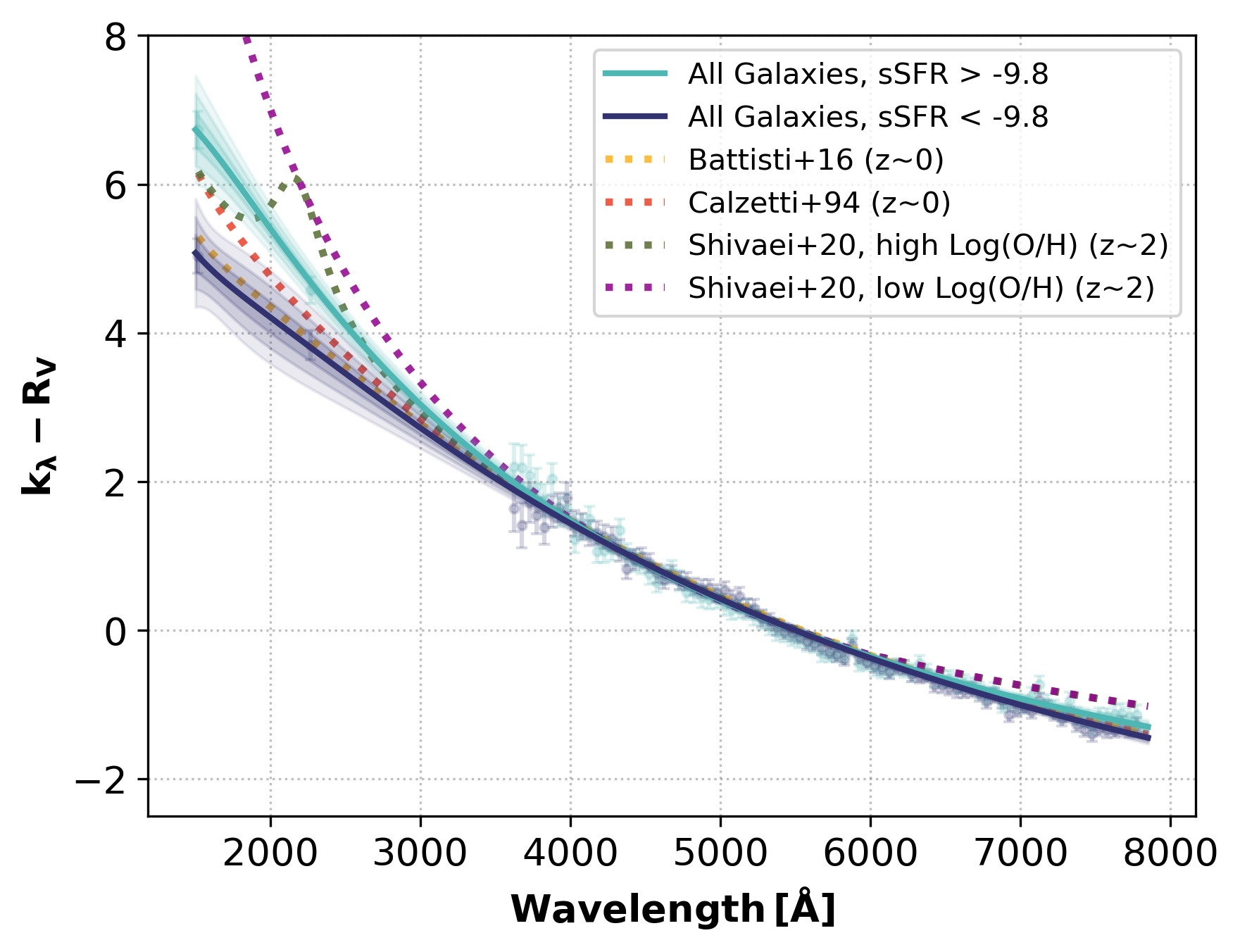} 
\caption{Attenuation curves binned as a function of the specific star formation rate. We split our sample at a Log(sSFR) of -9.8. We find that galaxies in the high-sSFR bin are associated with a steeper average attenuation curve than galaxies in the low-sSFR bin.}
\label{fig:tlbEvo}
\end{figure}

Whether a galaxy's specific star formation rate is correlated with the attenuation curve slope is a subject of ongoing debate. As both star formation rate and stellar mass are positively correlated with obscuration, and much recent work points to total obscuration as the dominant driver of variation in the attenuation curve \citep{Shivaei25,Salim18}, one might expect no correlation between curve shape and sSFR. In the local universe, some studies have found no evidence or only weak evidence \citep{Wild2011, Battisti2016} of variation as a function of sSFR. Meanwhile, other studies (particularly those using model-based rather than pairwise approaches) present a more complicated picture; rather than a monotonic trend, \citet{Salim18} finds that the curve steepens as one moves off from the center of the star-forming main sequence -- a particularly interesting result, as they find this trend is independent of the general dependence on dust opacity. Moving outside of the local universe, some work finds no relationship between sSFR and slope \citep{reddy15}, while other studies find a moderate link \citep{Markov25}, citing the lower average masses and metallicities of young starbursts as a potential explanation for steeper slopes in those systems. 

We look for such a relationship in our sample by splitting our sample at its median log(sSFR) = -9.8. The resulting curve is shown in Figure 11. We find a picture most consistent with results of \citet{Salim18}; there is a clear relationship between the slope of the attenuation curve and the sSFR, despite the sSFR being uncorrelated with the nebular opacity (as traced by $\tau_b$; Fig. 5c). In our sample, galaxies with higher specific star formation rates are clearly associated with a steeper attenuation curve. Both curves have values of $f$ and $R_V$ consistent with the \citep{calzetti} curve.

With the information we have available, it is difficult to definitively pinpoint what physically causes us to observe a steeper attenuation curve at high specific star formation rate. From a geometric perspective, one possibility is that systems with high specific star formation rate have star formation that is more concentrated (in a nuclear starburst, for example) compared to systems with less aggressive star formation. In such a scenario, less UV light would be scattered into the beam by dust when star formation is very concentrated in a single small region, resulting in a steeper attenuation curve. Grain-level effects may also play a role. It is well established that shocks from supernovae can destroy grains \citep{grains}. In extinction, small grains are associated with steeper extinction curves \citep[e.g.][and references therein]{asano14}. However, it is not clear how geometric and grain-level effects combine to produce the behavior we observe.

\section{Discussion}

\subsection{Nebular Opacity Can't Explain Curve Shapes}

\begin{figure}[ht!]
\gridline{\fig{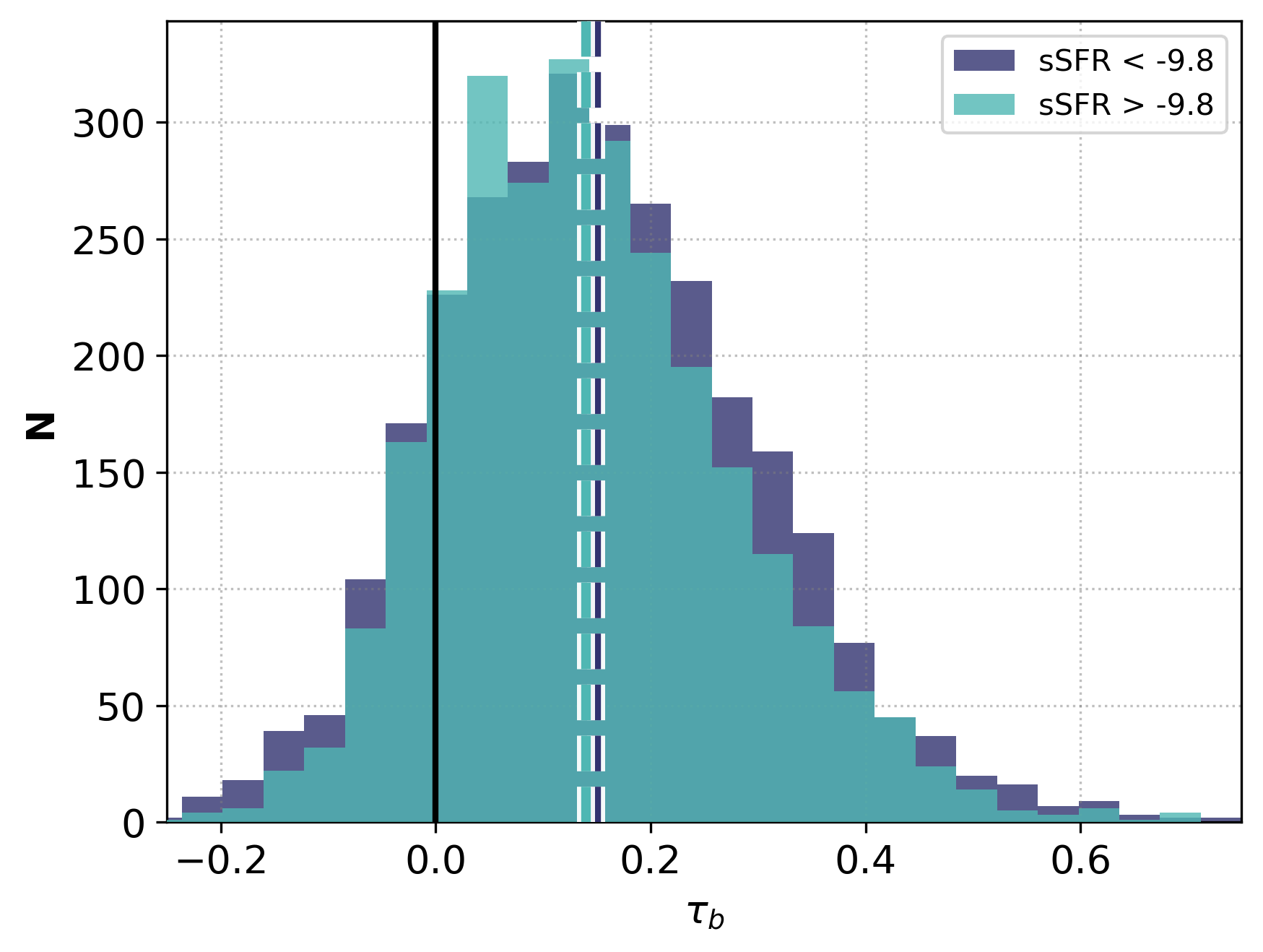}{0.98\columnwidth}{(a)}}
\vspace{-2mm}
\gridline{\fig{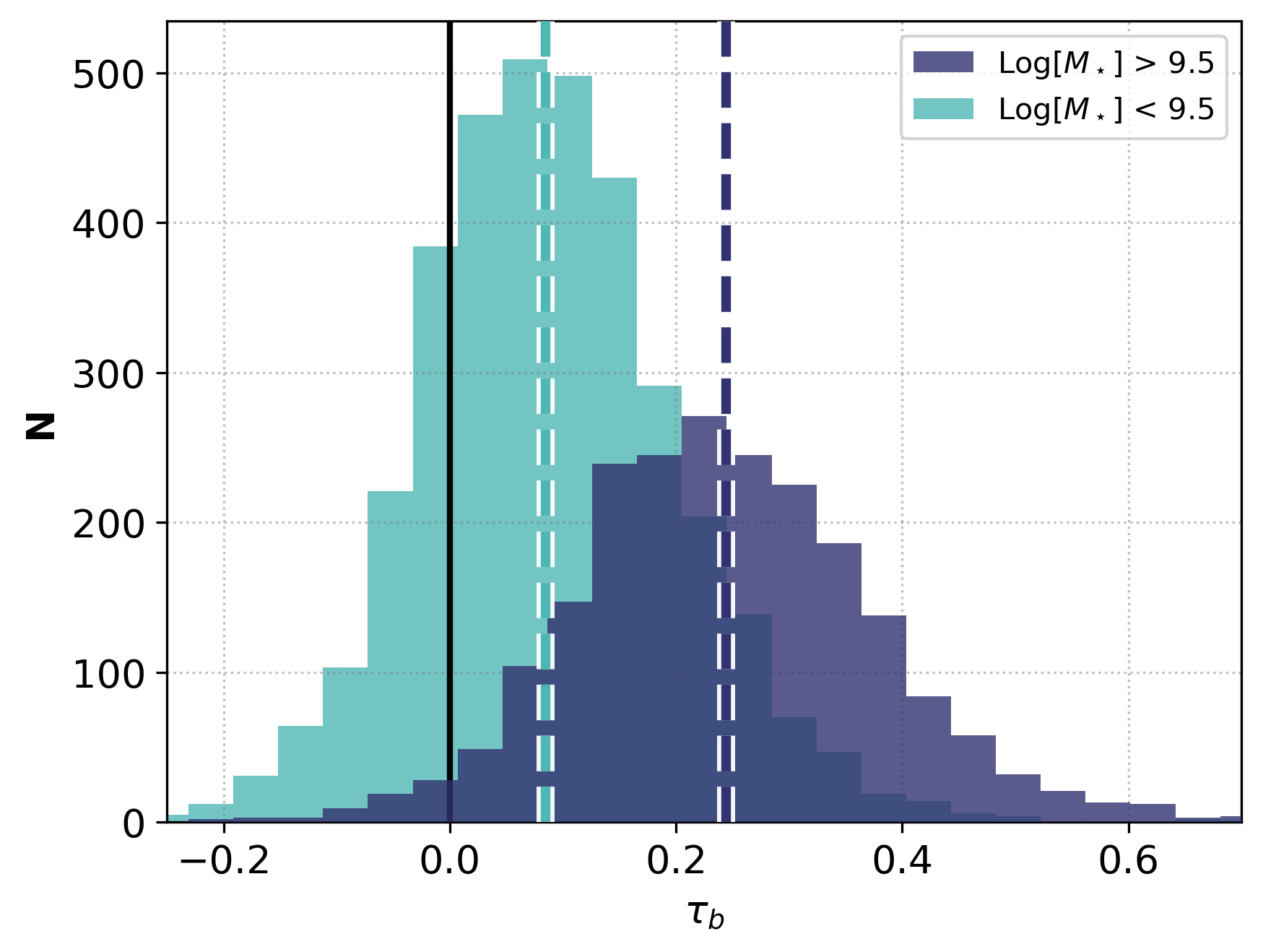}{0.98\columnwidth}{(b)}}
\caption{Histograms of $\tau_b$ as a function of sSFR (a) and stellar mass (b) for galaxies in our sample. The distributions have a nearly identical medians in (a), despite the sSFR driving significant variation in the curve slope. At the same time, the distributions in (b) are quite different despite our results showing minimal evolution as a function of stellar mass. This implies that the nebular opacity $\tau_b$ is not a primary driver of the attenuation curve slope. We note, however, that this does not imply that the total opacity $A_V$ does not impact curve slopes.} 
\label{fig:balmerDists}
\end{figure}
One weakness of this study is that we do not have reliable measurements of stellar continuum $A_V$ for the galaxies in our sample. As total dust opacity is considered a primary driver of variations in curve slope \citep[e.g.][]{Shivaei25,Salim18}, one may be tempted to use an alternative like $\tau_b$ as a proxy for $A_V$ when interpreting our results; however, such an assumption may not be well-justified.

Consider, for example, the curves we make as a function of stellar mass and specific star formation rate (Figures 8 and 10). There is obvious variation as a function of specific star formation rate (which is uncorrelated with $\tau_b$), while the relationship between stellar mass (which is correlated with $\tau_b$) and curve slope is less clear. A cut on stellar mass produces $\tau_b$ distributions with clearly different medians; simultaneously, there is no evidence that a cut on sSFR does the same (Figure 12). It is reasonable to infer from our results that the variation in the slope of the attenuation curve is not necessarily correlated with nebular opacity ($\tau_b$). It should also inspire confidence that the slope evolution we infer as a function of sSFR is not simply driven by a $\tau_b$ selection effect. 

We emphasize that this does not mean that our results necessarily contradict the $A_V$-driven picture of slope evolution. Total obscuration and Balmer optical depth are correlated, but loosely \citep{Salim18}, and there is evidence that the correlation between $\tau_b$ and stellar reddening (as traced by the UV slope $\beta$) is not particularly strong in our data (Figure 5). We are unable to draw direct conclusions about how opacity-related effects influence the variation in attenuation curve slopes we observe.


\subsection{Do We See A 2175$\AA$ Bump?}
With only a single filter (GALEX NUV) covering the 2175$\AA$ feature, it is difficult to draw conclusions about the presence or lack of such a bump in our data. As discussed in \citet{Battisti2016}, the presence of this feature in the GALEX NUV band may make the resulting attenuation curve shallower (resulting in bluer apparent UV spectra). With only a single data point on each curve sampling the feature, however, it is difficult to come to any strong conclusions. It might be that the relatively 'flat' appearance of our mass-divided attenuation curves (Figure 9) is due (at least in part) to the presence of a bump which is not accounted for in our fitting routine. However, it is unclear why binning galaxies as a function of stellar mass would make this feature stand out when it is not notable in our other curves; it is more likely that this behavior is just scatter.

\subsection{Future Work}
One weakness of this study is that our sample is not particularly diverse in $\tau_b$; there are few sources with a high level of nebular attenuation. Thus, it may be useful to bin galaxy stacks using a parameter other than $\tau_b$. Though stellar continuum $A_V$ would be the most physically meaningful choice, doing so would rely on assumptions about the shape of the attenuation curve and so may not be advisable. Instead, we are curious as to whether stacking as a function of $\beta_{obs}$ would produce reasonable attenuation curves. $\beta_{obs}$ is accessible through photometry at higher redshifts and might be a more direct tracer of the stellar reddening rather than the nebular reddening. However, as the intrinsic UV slope is also strongly affected by star formation history, we would need to restrict ourselves to galaxies of very similar age. The curves would then rely on the assumption that the intrinsic UV slopes are constant. It is unclear whether such an assumption would introduce more scatter than the set of assumptions attached to a $\tau_b$-anchored stack. It might also be interesting to do this exercise based on a metric derived from dust emission, like IRX (i.e. $L_{IR}/L_{UV}$). 

Future DESI data releases will include significantly larger galaxy samples. This will allow for a more robust analysis of what parameters actually drive change in the curve slope. In particular, if a sufficiently large sample is available, it might be possible to produce curves in two widely separated bins of, e.g., mass or metallicity, rather than splitting a continuous distribution in two at a single point. If the reason that we see minimal slope evolution as a function of mass, metallicity, and ionization parameter is that the bins we compare are too similar, this might make differences driven by these parameters more obvious.  

The large PSF and relatively poor sensitivity of GALEX also presented a challenge; scatter on the UV end of our attenuation curves was difficult to manage and made interpretation challenging. Future facilities like UVEX \citep{uvex} will dramatically increase both the quantity and quality of publicly available UV photometry, and in turn will likely dramatically advance our understanding of the attenuation curve in the local universe.

Perhaps most importantly, the galaxy sample used in this work might be well-suited to determine whether the different methods used to produce attenuation curves produce different average results. Here, we produced curves by comparing empirical galaxy SEDs; it is increasingly common to instead rely on SED modeling to do this \citep[e.g.][]{Shivaei25,Salim18}. We are very interested to apply the model method to these galaxies. It is not yet clear whether a systematic difference exists between attenuation curves derived using model-based methods and attenuation curves derived from stacks. Given that curves constructed using both methods are often compared directly in the literature, such an experiment may assist with interpretation.

\section{Summary and Conclusions}

In this work we use a sample of 3457 star-forming galaxies drawn from DESI DR1 and cross-matched to GALEX FUV/NUV photometry to construct new UV–optical attenuation curves at $z\simeq0$–0.25 via the pairwise method. Our main conclusions are as follows:

\begin{enumerate}
    \item We find a positive linear correlation between the Balmer optical depth $\tau_b$ and the UV slope $\beta$. The slope of the relationship between these two parameters is consistent with existing work.

    \item We derive attenuation curves by comparing stacked spectral templates binned as a function of $\tau_b$. We construct a single average 'bulk' attenuation curve containing all galaxies in our sample, as well as multiple additional curves produced by splitting that sample as a function of mass, metallicity, ionization parameter, and sSFR. The `bulk' selective attenuation curve we derived is consistent with previous low-redshift results \citep[e.g.][]{Calzetti94,Battisti2016}.  

    \item The slope of the attenuation curve appears to depend strongly on the specific star formation rate, such that a higher specific star formation rate is associated with a steeper attenuation curve. It is not clear what causes this behavior, but both geometric (related to the location of star formation relative to dust) and grain-level effects (related to the grain size distribution and how it can be modified by feedback) may play a role. We find that there is no clear link between the slope of the attenuation curve and stellar mass, gas-phase metallicity, and ionization parameter.
    
    \item Galaxy-galaxy variation in the steepness of the attenuation curve cannot be explained by differences in the amount of nebular dust opacity. However, we can neither verify opacity-independent variation nor verify which of these (inter-correlated) parameters are responsible for slope variation without producing samples that are matched in every parameter except for the parameter of interest. In the future, larger galaxy samples may make this possible. 


\end{enumerate}


\begin{acknowledgments}

\begingroup
\footnotesize
A.M. thanks Irene Shivaei for insightful discussion.
Part of this work has been supported by NASA, via the Jet
Propulsion Laboratory Euclid Project Office, as part of the
“Science Investigations as Members of the Euclid Consortium
and Euclid Science Team” program.
This work is based on observations made with the NASA
Galaxy Evolution Explorer. GALEX is operated for NASA by
the California Institute of Technology under NASA contract
NAS5-98034.
This research uses services or data provided by the SPectra Analysis and Retrievable Catalog Lab (SPARCL), which is part of the Community Science and Data Center (CSDC) program at NSF National Optical-Infrared Astronomy Research Laboratory. NOIRLab is operated by the Association of Universities for Research in Astronomy (AURA), Inc. under a cooperative agreement with the National Science Foundation.
This material is based upon work supported by the U.S. Department of Energy (DOE),Office of Science, Office of High-Energy Physics, under Contract No. DE-AC02-05CH11231,and by the National Energy Research Scientific Computing Center, a DOE Office of Science User Facility under the same contract. Additional support for DESI was provided by theU.S. National Science Foundation (NSF), Division of Astronomical Sciences under Contract No. AST-0950945 to the NSF’s National Optical-Infrared Astronomy Research Laboratory;the Science and Technology Facilities Council of the United Kingdom; the Gordon and Betty Moore Foundation; the Heising-Simons Foundation; the French Alternative Energies and Atomic Energy Commission (CEA); the National Council of Humanities, Science and Technology of Mexico (CONAHCYT); the Ministry of Science, Innovation and Universities of Spain (MICIU/AEI/10.13039/501100011033), and by the DESI Member Institutions: https://www.desi.lbl.gov/collaborating-institutions. Any opinions, findings, and conclusions or recommendations expressed in this material are those of the author(s) and do not necessarily reflect the views of the U.S. National Science Foundation, the U.S. Department of Energy, or any of the listed funding agencies. This material is based upon work supported by the U.S. Department of Energy, Officeof Science, Office of Workforce Development for Teachers and Scientists, Office of Science Graduate Student Research (SCGSR) program. The SCGSR program is administered by the Oak Ridge Institute for Science and Education (ORISE) for the DOE. ORISE is managed by ORAU under contract number DESC0014664. All opinions expressed in this paper are the author’s and do not necessarily reflect the policies and views of DOE, ORAU, or ORISE. The DESI Legacy Imaging Surveys consist of three individual and complementary projects:the Dark Energy Camera Legacy Survey (DECaLS), the Beijing-Arizona Sky Survey (BASS),and the Mayall z-band Legacy Survey (MzLS). DECaLS, BASS and MzLS together include data obtained, respectively, at the Blanco telescope, Cerro Tololo Inter-American Observatory, NSF’s NOIRLab; the Bok telescope, Steward Observatory, University of Arizona; and the Mayall telescope, Kitt Peak National Observatory, NOIRLab. NOIRLab is operated by the Association of Universities for Research in Astronomy (AURA) under a cooperative agreement with the National Science Foundation. Pipeline processing and analyses of the data were supported by NOIRLab and the Lawrence Berkeley National Laboratory. Legacy Surveys also uses data products from the Near-Earth Object Wide-field Infrared Survey Explorer(NEOWISE), a project of the Jet Propulsion Laboratory/California Institute of Technology,funded by the National Aeronautics and Space Administration. Legacy Surveys was supported by: the Director, Office of Science, Office of High Energy Physics of the U.S. Department of Energy; the National Energy Research Scientific Computing Center, a DOE Office of Science User Facility; the U.S. National Science Foundation, Division of Astronomical Sciences; the National Astronomical Observatories of China, the Chinese Academy of Sciences and the Chinese National Natural Science Foundation. LBNL is managed by the Regents of the University of California under contract to the U.S. Department of Energy. The complete acknowledgments can be found at https://www.legacysurvey.org/. Any opinions, findings, and conclusions or recommendations expressed in this material are those of the author(s) and do not necessarily reflect the views of the U.S. National Science Foundation, the U.S. Department of Energy, or any of the listed funding agencies. The authors are honored to be permitted to conduct scientific research on I’oligam Du’ag(Kitt Peak), a mountain with particular significance to the Tohono O’odham Nation.
\endgroup

\end{acknowledgments}

\appendix
\newpage
\section{How Does Our Galaxy Sample Compare To The DESI Parent Sample?}
Making reliable attenuation curves required us to apply a large number of cuts to the DESI and GALEX parent samples as described in Section 2. Both the cuts we explicitly apply as well as the effective cut applied by cross-matching DESI and GALEX massively reduce the number of galaxies relative to each parent sample. As a result, the galaxy population from which we measure the attenuation curve is not necessarily representative of the DESI sample as a whole. Most importantly, the cuts we perform result in a significantly less massive, less obscured galaxy population (Figure 14). Both the explicit cuts and the cross-match with GALEX contribute to this shift.

\begin{figure*}[htb!]
  \centering
  \includegraphics[width=0.49\textwidth]{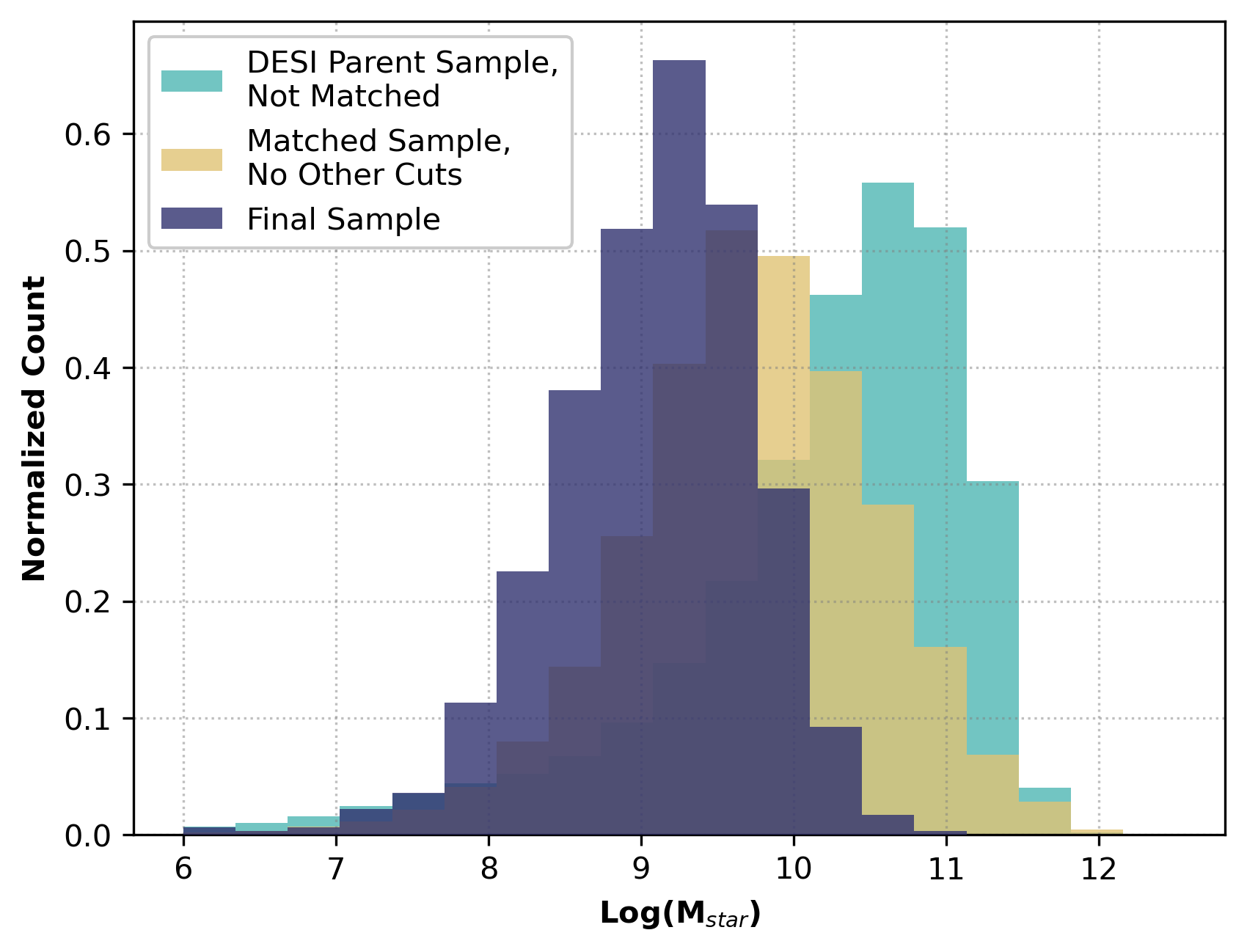}\hfill
  \includegraphics[width=0.49\textwidth]{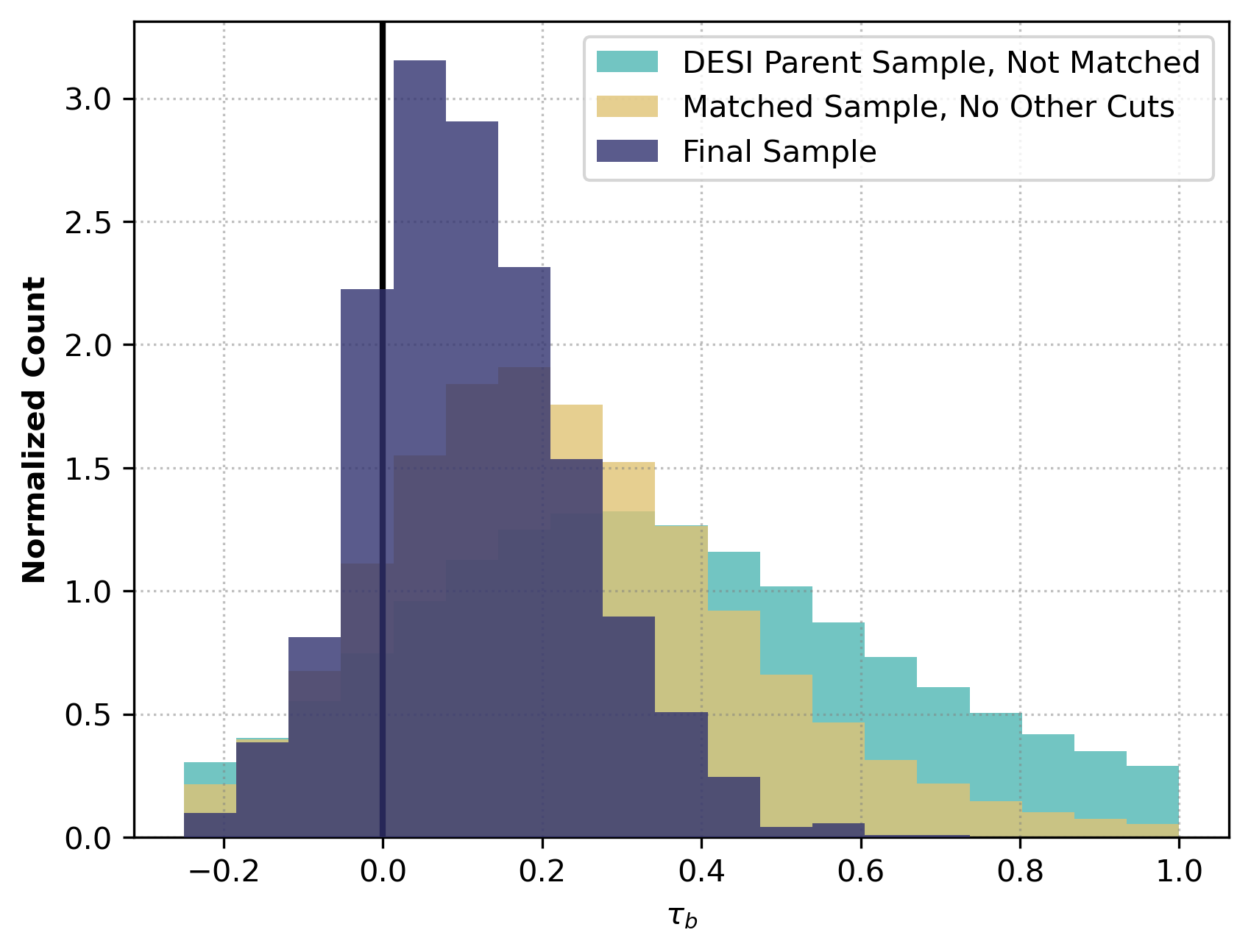}
  \caption{The mass (left) and $\tau_b$ (right) distributions of the DESI parent sample (light blue), that sample after cross-matching with GALEX but with no other cuts applied (beige), and after the remaining quality cuts (dark blue). Both our quality cuts and the cross-matching serve to select galaxies with lower masses and Balmer optical depths on average.}
  \label{fig:comp}
\end{figure*}

\newpage
\section{Sensitivity Of Metallicity Evolution On Calibrations}
A vast array of different metallicity diagnostics are available in the literature. In this work, we settled on the use of the \citep{o3n2} O3N2 diagnostic. However, arguably more advanced metallicity diagnostics exist, such as the S-cal developed in \citet{Pilyugin16}. Though this diagnostic may produce more reliable metallicities for individual galaxies, it comes at the disadvantage of requiring a large number of emission lines to be used. We find that making the line SNR cuts necessary to use the S-cal diagnostic reduced the size of our sample by a factor of about 3. We found that splitting the sample in two as a function of metallicity, as determined by this diagnostic, also did not produce significantly different attenuation curves. However, the curves are significantly less reliable due to the reduced sample size -- particularly in the low-metallicity case. As such, we decided to use a metallicity tracer that allows us to retain our entire sample.

\begin{figure}[htb!]
\centering
\includegraphics[width=0.49\textwidth]{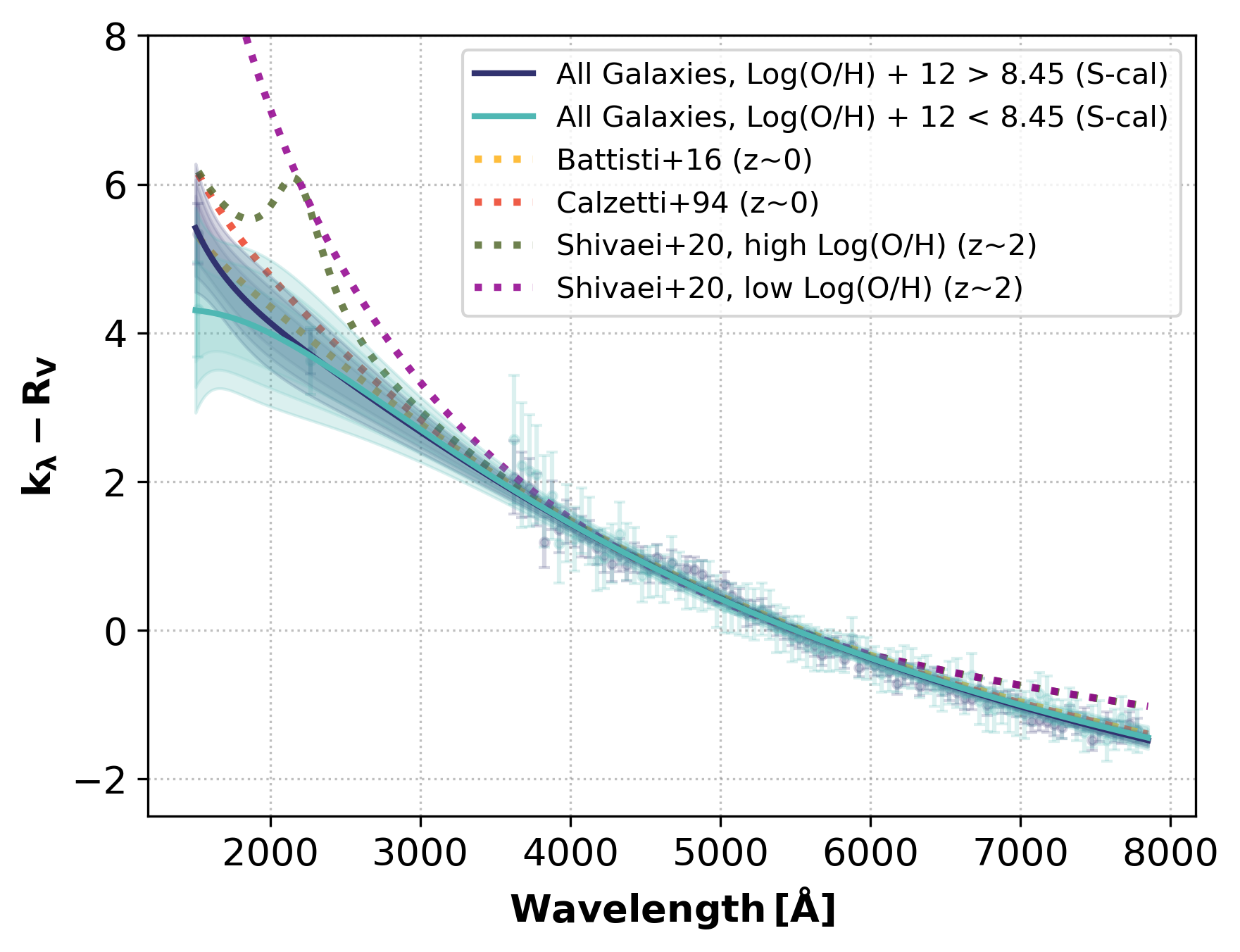} 
\caption{Metallicity-binned attenuation curves split at Log(O/H) + 12 = 8.45, using the S-cal diagnostic. As before, we find no significant difference in shape between our low-metallicity and high-metallicity attenuation curves. However, the decreased sample size drastically increases the uncertainty inherent to these curves, particularly in the low-metallicity case where strong scatter in the UV is readily apparent. In this plot, 719 galaxies were used in the high-Z curve and 574 were used in the low-Z curve.}
\label{fig:metals}
\end{figure}

\newpage
\section{Redshift Dependence Of The Attenuation Curve}
To test whether it is reasonable to treat our z=0.0-z=0.25 sample as a single redshift bin, we divide the sample into two bins at z=0.125 and produce attenuation curves within each subsample. We find that there is no significant difference between the curves at low and high redshift, which demonstrates that our approach is reasonable. This result does not change if selections based on, e.g., ensuring the same mass completeness in all redshift bins are adopted.

\begin{figure*}[htb!]
  \centering
  \includegraphics[width=0.49\textwidth]{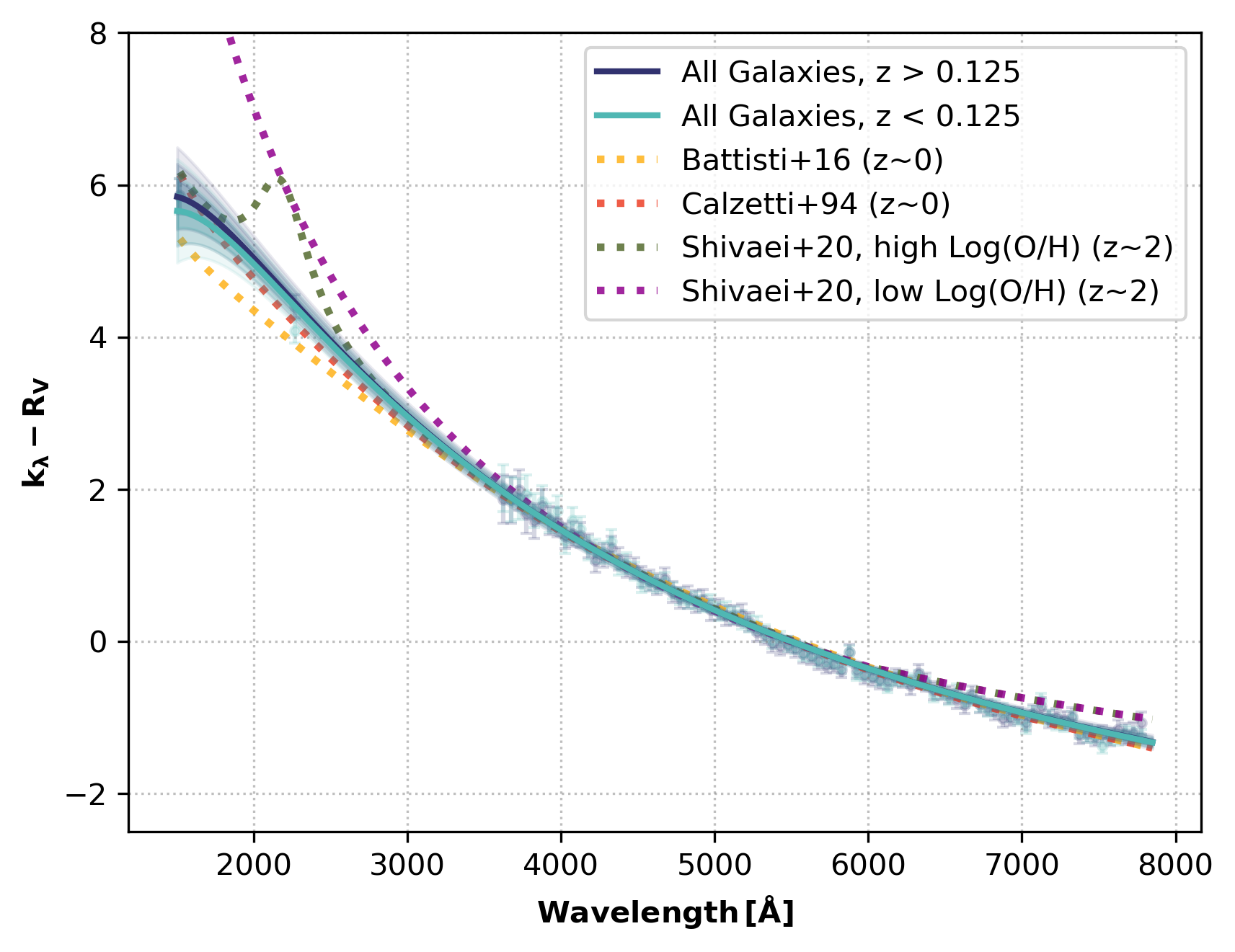}\hfill
  \caption{Redshift-binned attenuation curves split at z=0.125 with no further cuts applied.
  The curves are consistent with one another across redshift. This demonstrates that it is not unreasonable for us to treat our sample as a single redshift bin, and shows that there is not strong evidence for redshift evolution in our sample.}
  \label{fig:z}
\end{figure*}




\clearpage
\bibliography{sample7}{}
\bibliographystyle{aasjournalv7}

\end{document}